\documentclass[11pt]{article}
\usepackage{amssymb,amsmath}
\usepackage{ascmac}
\usepackage[dvipdfmx]{hyperref}
\usepackage[dvips]{graphicx,psfrag}
\usepackage{epsfig}
\usepackage{array}
\usepackage{lscape}
\usepackage{accents}
\usepackage{color}
\usepackage{braket}
\usepackage{bm}
\usepackage{multirow}

\numberwithin{equation}{section}

\makeatletter
\def\alt{\mathrel{\mathpalette\gl@align<}}
\def\agt{\mathrel{\mathpalette\gl@align>}}
\def\gl@align#1#2{\lower.6ex\vbox{\baselineskip\z@skip\lineskip\z@
\ialign{$\m@th#1\hfil##\hfil$\crcr#2\crcr\sim\crcr}}}
\makeatother

\newcommand{\doi}[1]{\href{https://doi.org/#1}{doi:#1}}

\newcommand{\mubar}{\overline{\text{Mu}}}
\newcommand{\dash}{\hspace{1pt}--\hspace{1pt}}

\begin{document}

\begin{flushright}
{\large
June, 2022
}
\end{flushright}
\vspace*{5mm}

\begin{center}
\baselineskip 20pt

{\bf
\LARGE
Neutrinoless double beta decay 
and the muonium-to-antimuonium transition 
in models \\ 
with a doubly charged scalar
} 

\vspace{1.0cm}

{\Large 
Takeshi Fukuyama$\,{}^{a}$,
Yukihiro Mimura$\,{}^{b}$
and
Yuichi Uesaka$\,{}^{c}$
}

\vspace{8mm}

{\large
${}^a${\it 
Research Center for Nuclear Physics (RCNP),
Osaka University, \\Ibaraki, Osaka, 567-0047, Japan
}\\
\vspace{3mm}
${}^{b}${\it
Department of Physical Sciences, College of Science and Engineering, \\
Ritsumeikan University, Shiga 525-8577, Japan
}\\
\vspace{3mm}
${}^{c}${\it
Faculty of Science and Engineering, Kyushu Sangyo University, \\
2-3-1 Matsukadai, Higashi-ku, Fukuoka 813-8503, Japan
}\\
}

\vspace{1.2cm}

{\Large
{\bf Abstract}}\end{center}
\baselineskip 18pt
{\large
The lepton number and flavor violations are important possible ingredients 
of the lepton physics.
The neutrinoless double beta decay and the transition of the muonium 
into antimuonuim are related to those violations.
The former can give us an essential part of fundamental physics,
and there are plenty of experimental attempts to observe the process.
The latter has also been one of the attractive phenomena,
and the experimental bound will be updated in planned experiments
at new high-intensity muon beamlines.
In models with a doubly charged scalar,
not only can those two processes be induced,
but also the active neutrino masses can be induced radiatively.
The flavor violating decays of the charged leptons 
constrain the flavor parameters of the models.
We study how 
the muonium-to-antimuonium transition rate can be
as large as the current experimental bound,
and we insist that the updated bound of the transition rate 
will be useful to distinguish the models to generate the neutrinoless double beta decay 
via the doubly charged scalar.
We also study a possible extension of the model to the left-right model
which can induce the neutrinoless double beta decay.
}


\thispagestyle{empty}
\newpage
\addtocounter{page}{-1}

\section{Introduction}
\label{sec1}

\fontsize{13pt}{17pt}\selectfont

The neutrinoless double beta decay $(0\nu 2\beta$ decay) is one of the key phenomena
to probe lepton number violation.
The experimental searches for the $0\nu 2\beta$ decay 
from $^{76}$Ge \cite{Agostini:2020xta},
$^{130}$Te \cite{CUORE:2021mvw}, and 
 $^{136}$Xe \cite{KamLAND-Zen:2016pfg} are
ongoing, and the $0\nu 2\beta$ half-value periods of them are bounded to be more than about $10^{25}$\dash$10^{26}$ years depending on the nuclei.
The next-generation experiments including other isotopes are 
planned \cite{LEGEND:2021bnm,SNO:2021xpa,nEXO:2021ujk,CUPID:2019imh,CANDLES:2020iya,Dolinski:2019nrj,Agostini:2022zub},
and hundred times higher sensitivity of the half-lives is targeted.
Together with precise data from the neutrino oscillation experiments and cosmological observations
of the sum of the neutrino masses, 
it is expected to advance our understanding of the mass structure of neutrinos.

On the lepton physics, 
the high intensity muon beamline is being upgraded \cite{Kawamura:2018apy,Aiba:2021bxe}, 
and precise measurements of the muon properties are planned.
The transition of the muonium $(\mu^+ e^-)$ into antimuonium ($\mu^- e^+$) (Mu-to-$\mubar$ transition) 
\cite{Pontecorvo:1957cp,Feinberg:1961zza,Lee:1977tib} is 
one of the phenomena to search at the facilities.
It has been pointed that the transition and the $0\nu 2\beta$ decay can have a close relationship \cite{Halprin:1982wm}.
The Mu-to-$\mubar$ transition violates lepton flavor numbers ($\Delta L_\mu =  -\Delta L_e = 2$).
Although lepton flavor violation (LFV) can be easily considered in models beyond the standard model (SM),
the LFV decays, such as $\mu \to e\gamma$ \cite{Adam:2013mnn} and $\mu \to 3e$ \cite{Bellgardt:1987du},
and $\mu$\dash$e$ conversion in nuclei \cite{SINDRUMII:2006dvw} give severe constraints to the models.
Those LFV processes violate lepton flavor numbers with single units ($\Delta L_e, \Delta L_\mu = \pm 1$).
The Mu-to-$\mubar$ transition can be allowed theoretically even if the LFV decays of the charged leptons are severely suppressed.
The Mu-to-$\mubar$ transition experiment \cite{Willmann:1998gd} 
provides the bound of the coefficients of the four-fermion operators to induce the transition
($\sim 10^{-3}$ in the unit of the Fermi coupling constant $G_F$), and the bound will be lowered
to $ O(10^{-4}) G_F$ by the
experiments at the facilities with the upgraded muon beamlines  \cite{Kawamura:2021lqk,Han:2021nod,Bai:2022sxq}.

The (1,1) element of the neutrino mass matrix, $m_{ee}$, gives a contribution
of the $0\nu 2\beta$ amplitude from the ordinary diagram
with the long-range exchange of the Majorana neutrino.
If there are particles beyond the SM
and there is a new lepton number violating interaction, 
the $0\nu 2\beta$ amplitude can be generated \cite{Vergados:2002pv,Vergados:2012xy}
even if $m_{ee}$ is small.
The doubly charged scalar is one of the representative new particles
that can induce the $0\nu 2\beta$ decay via short-range interactions.
In the models with the doubly charged particle,
the active neutrino mass can be also induced radiatively
(see Ref.\cite{Cai:2017jrq} for a review of the radiative neutrino mass models).
The Majorana masses of the active neutrinos and the $0\nu 2\beta$ decay are
linked by the lepton number violation.
Various possibilities of such types of the models that can induce the $0\nu 2\beta$ decay 
have been considered \cite{Chen:2006vn,Petcov:2009zr,Gustafsson:2014vpa,King:2014uha,Liu:2016mpf,Alcaide:2017xoe,Geng:2020gsu,Chen:2021rcv}.
The doubly charged scalar couplings to the leptons can carry lepton flavor numbers,
and therefore,
the Mu-to-$\mubar$ transition can be also induced via the doubly charged scalar exchange \cite{Chang:1989uk,Swartz:1989qz}
in those models.
The Mu-to-$\mubar$ transition can be a tool to select models
to induce the $0\nu 2\beta$ decay,
which is what we focus on in this paper.

\begin{figure}
\center
\includegraphics[width=6cm]{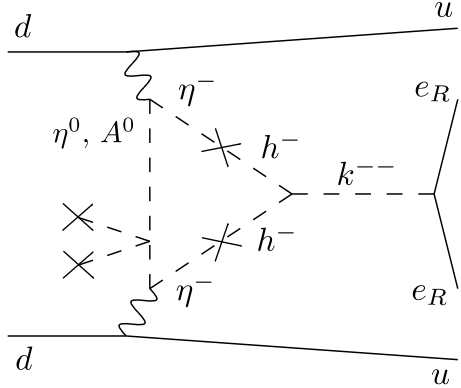}
\caption{
Cocktail diagram to induce the $0\nu 2\beta$ decay.
The $\times$ mark stands for the insertion of the electroweak vev of the Higgs boson.
The $h^-$ and $k^{--}$ fields are $SU(2)_L$ singlet scalars, and 
$\eta^-$, $\eta^0$ and $A^0$ are contained in the $SU(2)_L$ doublet scalar which does not
acquire a vev. 
}
\label{fig1}
\end{figure}

If $SU(2)_L$ singlet scalars, $h^+$ and $k^{++}$, with their hypercharges $Y=1$ and $Y=2$ are added
in the two-Higgs-doublet model (2HDM),
the $0\nu 2\beta$ decay can be induced via the diagram given in Fig.\ref{fig1}.
We call that diagram a cocktail diagram \cite{Gustafsson:2012vj}.
In the model, the charged scalar $\eta^-$ in the $SU(2)_L$ Higgs doublet is mixed with the singlet $h^+$,
and the mixing violates the lepton number.
The active neutrino masses can be induced at the two-loop level.
We call this model the Liu-Gu model since they discuss both the $0\nu 2\beta$ decay
and the loop-induced neutrino mass matrix in their model in Ref.\cite{Liu:2016mpf}.
The half-value period of the $0\nu 2\beta$ decay via the cocktail diagram can be 
as large as the current experimental bound.
The singlet $h^+$ can have bilinear couplings to the left-handed lepton doublets $\ell$,
and then, two-loop induced neutrino masses can be obtained.
The model with the $\ell \ell h^+$ coupling is called the Zee-Babu model \cite{Zee:1980ai,Zee:1985id,Babu:1988ki,Babu:2002uu}.
In the original Zee-Babu model, 
only one Higgs doublet is enough to generate the neutrino masses
and thus, the $0\nu 2\beta$ decay is induced by $m_{ee}$ from the ordinary diagram. 
As a consequence, the half-value period of the decay is much larger than the current bound
in the normal neutrino mass hierarchy.
Authors have shown that the Mu-to-$\mubar$ transition rate can be as large as
the current experimental bound in the Zee-Babu model \cite{Fukuyama:2021iyw}.
In the Liu-Gu model, on the other hand, 
the Mu-to-$\mubar$ transition rate is bounded to be much smaller than 
the current bound due to experimental constraints of the LFV decays of the charged leptons.
Therefore, the $0\nu 2\beta$ decay and the Mu-to-$\mubar$ transition
are useful to distinguish those two models.

Can both the $0\nu 2\beta$ decay amplitude and the Mu-to-$\mubar$ transition rate
be as large as the current bounds?
In this paper, we show that
the Mu-to-$\mubar$ transition rate can be as large as the current bound
avoiding the constraints from the LFV decays
and the $0\nu 2\beta$ decay is induced by the cocktail diagram in Fig.\ref{fig1}
if the $\ell \ell h^+$ coupling is added in the Liu-Gu model
(in other words, the $\eta^+$\dash$h^+$ mixing is simply added in the Zee-Babu model).
Those amplitudes are proportional to the $e_R e_R k^{++}$ coupling,
and the scalar spectrum is predictable to induce the half-value period of the $0\nu 2\beta$ decay
and the Mu-to-$\mubar$ transition.
In the Liu-Gu model, the type II (or type X) 2HDM is preferable 
to induce the neutrino masses with suppressing the LFV decays.
In order to induce the sizable Mu-to-$\mubar$ transition operator,
it is possible to apply 
the type I (or type Y) 2HDM in our hybrid model,
and the scalars in the doublets can be lighter
while satisfying the experimental constraints from their decays.
If the scalars in the doublets are lighter, the half-value period of the $0\nu 2\beta$ decay
can be shorter.
We will also describe the $0\nu 2\beta$ decay and the Mu-to-$\mubar$ transition
in the left-right model 
in which the $h^+$ and $k^{++}$ scalars can be unified in a $SU(2)_R$ triplet.

This paper is organized as follows:
In Section \ref{sec2}, we introduce the 2HDM with the $SU(2)_L$ singlet scalars $h^+$ and $k^{++}$,
and we present an overview of the Liu-Gu and Zee-Babu models.
In Section \ref{sec3}, we describe the $0\nu 2\beta$ decay amplitude via the cocktail diagram in Fig.\ref{fig1},
and show the numerical calculation of the half-value periods.
In Section \ref{sec4}, the loop-induced neutrino masses are given.
In Section \ref{sec5}, the Mu-to-$\mubar$ transition in Liu-Gu and Zee-Babu models are discussed.
In Section \ref{sec6}, we exhibit the analyses for our model that can make both amplitudes of
the $0\nu 2\beta$ decay and Mu-to-$\mubar$ transition as large as the current experimental bounds,
and discuss how the experimental data can distinguish the models.
In Section \ref{sec7}, we develop the model to the left-right model with a $SU(2)_R$ triplet.
Section \ref{sec8} gives the conclusion and discussion of this paper.
In Appendix \ref{appendix:A}, the scalar spectrum in the 2HDM with the $h^+$ and $k^{++}$ scalars are given.
The condition to avoid the charge symmetry breaking vacua is also noted.
In Appendix \ref{appendix:B}, the loop functions for the cocktail and box diagrams are given.
In Appendix \ref{appendix:C}, the loop functions for the neutrino masses are shown.
In Appendix \ref{appendix:D}, the LFV constraints are summarized.
In Appendix \ref{appendix:E}, the scalar spectrum in the left-right model with a $SU(2)_R$ triplet is given.

\section{Overview}
\label{sec2}

We introduce two $SU(2)_L$ doublets, $\Phi_1$ and $\Phi_2$, and two $SU(2)_L$ singlets, $h^+$ and $k^{++}$, with hypercharges $Y=1$ and $Y=2$, respectively.
The doublets acquire vacuum expectation values (vevs)
to break $SU(2)_L \times U(1)_Y$:
\begin{equation}
\langle \Phi_1 \rangle = \left(
 \begin{array}{c}
  v_1 \\ 0
 \end{array}
\right),
\qquad
\langle \Phi_2 \rangle = \left(
 \begin{array}{c}
  v_2 \\ 0
 \end{array}
\right).
\end{equation}
We define
\begin{equation}
\tan\beta = \frac{v_2}{v_1},
\end{equation}
and
\begin{align}
 \Phi &=  c_\beta \Phi_1 + s_\beta \Phi_2, \\
 \eta &=  -s_\beta \Phi_1 +c_\beta \Phi_2,
\end{align}
with a notation: $c_\beta = \cos\beta$ and $s_\beta = \sin\beta$,
so that the vev of the doublet $\eta$ is zero.
The components in $\Phi$ and $\eta$
are
\begin{equation}
\Phi = 
\left(
 \begin{array}{c}
  v + \frac{\Phi^0 + i \omega^0}{\sqrt2} \\ \omega^-
 \end{array}
\right),
\qquad
\eta  = \left(
 \begin{array}{c}
   \frac{\eta^0 + i A^0}{\sqrt2} \\ \eta^-
 \end{array}
\right),
\end{equation}
where $v= \sqrt{v_1^2 + v_2^2}$, and $\omega^-$ and $\omega^0$ are the would-be Nambu-Goldstone bosons
eaten by the $W$ and $Z$ gauge bosons.

To avoid flavor changing neutral currents,
a discrete symmetry is often assumed in the 2HDM \cite{Haber:1984rc,Gunion:1989we}.
For the Yukawa coupling of the up-type quarks, the one with $\Phi_2$ is conventionally chosen.
For the Yukawa couplings of the down-type quarks and the charged leptons,
four types of choices are possible and they are referred to as type I, II, X, and Y.
Leaving out the selection for the down-type quarks,
Table \ref{Table1} shows the choices of the discrete symmetry for the charged leptons.
The coupling of the charged scalar in $\eta$ to the leptons are
\begin{align}
-{\cal L} &\supset - \frac{(M_e)_{ij}}{v} \cot\beta \, \overline{e_{iR}} \nu_{jL} \eta^-  \quad \mbox{in type I/Y}, \\
-{\cal L} &\supset + \frac{(M_e)_{ij}}{v} \tan\beta \, \overline{e_{iR}} \nu_{jL} \eta^-  \quad \mbox{in type II/X} ,
\end{align}
where $M_e$ is the charged lepton mass matrix, $M_e = {\rm diag} ( m_e, m_\mu, m_\tau)$.

The lepton couplings to $h^+$ and $k^{++}$ are given as follows:
\begin{equation}
{\cal L} = f_{ij} \overline{\ell^c_i} \ell_j h^+ + g_{ij} \overline{(e_{iR})^c} e_{jR} k^{++} + {\rm H.c.},
\label{f-and-g}
\end{equation}
where $\ell$ denotes the left-handed lepton doublets, $\ell = (\nu_L,e_L)^T$,
and
$\overline{\ell^c_i} \ell_j$ (with a proper $SU(2)_L$ contraction) can be written as
\begin{equation}
\overline{\ell^c_i} \ell_j = \overline{(\nu_{iL})^c} e_{jL} - \overline{(e_L)^c_i} \nu_{jL} =  \overline{(\nu_{iL})^c} e_{jL} - \overline{(\nu_{jL})^c} e_{iL}.
\end{equation}
The coupling $f_{ij}$ is anti-symmetric under the flavor indices, while the coupling $g_{ij}$ is symmetric.

The scalar potential in terms of $\Phi_1$, $\Phi_2$, $h^+$, and $k^{++}$ is
\begin{align}
V &= m_1^2 \Phi_1^\dagger \Phi_1 + m_2^2 \Phi_2^\dagger \Phi_2 - m_3^2 (\Phi_1^\dagger \Phi_2 + {\rm H.c.})
+ \frac{\lambda_1}2 (\Phi_1^\dagger \Phi_1)^2 
+ \frac{\lambda_2}2 (\Phi_2^\dagger \Phi_2)^2  \nonumber \\
&+
\lambda_3 \Phi_1^\dagger \Phi_1\Phi_2^\dagger \Phi_2
+
\lambda_4 \Phi_1^\dagger \Phi_2\Phi_2^\dagger \Phi_1
+
\frac{\lambda_5}{2} ((\Phi_1^\dagger \Phi_2)^2 + {\rm H.c.}) \nonumber\\
&+
(m_h^{\prime  2} + \lambda_{h\Phi_1} \Phi_1^\dagger \Phi_1 + \lambda_{h\Phi_2} \Phi_2^\dagger \Phi_2  ) |h^+|^2
+
(m_k^{\prime  2} + \lambda_{k\Phi_1} \Phi_1^\dagger \Phi_1 + \lambda_{k\Phi_2} \Phi_2^\dagger \Phi_2  ) |k^{++}|^2 \nonumber\\
&+
\lambda_h |h^+|^4 + \lambda_k |k^{++}|^4 + \lambda_{hk} |h^+|^2 |k^{++}|^2 
+ (\lambda_{hk\Phi\Phi} h^- k^{++} \Phi_1 \Phi_2+ {\rm H.c.})\nonumber
\\
&+
(m_{hhk} h^- h^- k^{++} + m_{\Phi \Phi h}   \Phi_1 \Phi_2 h^+ + {\rm H.c.}),
\label{scalar-potential}
\end{align}
where only those allowed by the $Z_2$ symmetry are written 
for the quartic couplings in terms of $\Phi_1$ and $\Phi_2$.
We define the vev-dependent masses of $h^+$ and $k^{++}$ as
\begin{equation}
m_{h^+}^2 = m_h^{\prime  2} + \lambda_{h\Phi_1} v_1^2 + \lambda_{h\Phi_2} v_2^2,
\qquad
m_{k}^2 = m_k^{\prime  2} + \lambda_{k\Phi_1} v_1^2 + \lambda_{k\Phi_2} v_2^2.
\end{equation}
The neutral scalars $\Phi^0$ and $\eta^0$ are mixed (with the mixing angle $\beta- \alpha$),
and their mass eigenstates are denoted by $h^0$ and $H^0$.
The masses of the neutral scalars, $h^0$, $H^0$, and $A^0$, are denoted as
$m_h, m_H$, and $m_A$.
The SM(-like) Higgs boson is $h^0$, and $m_h = 125$ GeV.
The singly charged scalars, $\eta^+$ and $h^+$, can be mixed by the $\Phi_1 \Phi_2 h^+ = \Phi \eta h^+$ term,
and we denote the mass eigenstates as $h_1^+$ and $h_2^+$,
and their masses as
$m_{h_1^+}$ and $m_{h_2^+}$.
We denote the mixing angle of them as $\phi$, and
\begin{equation}
\left( 
 \begin{array}{c}
  h_1^+ \\ h_2^+ 
 \end{array}
\right)
= 
\left( 
 \begin{array}{cc}
  c_\phi & -s_\phi \\
  s_\phi& c_\phi
 \end{array}
\right)
\left( 
 \begin{array}{c}
  \eta^+ \\ h^+ 
 \end{array}
\right),
\qquad
\tan2\phi = \frac{2v m_{\Phi\Phi h}}{m_{h^+}^2 - m_{\eta^+}^2}.
\end{equation}
The expressions of the scalar masses are given in Appendix \ref{appendix:A}.

\begin{table}[t]
\caption{$Z_2$-charge assignments for leptons and scalars.
}
\center
\begin{tabular}{c|c|cccccc}
\hline
\multicolumn{2}{c|}{}
& $\ell$  &  $e_R$ & $\Phi_1$ & $\Phi_2$ & $h^+$ & $k^{++}$ \\
\hline
\multicolumn{2}{c|}{$SU(2)_L \times U(1)_Y$} 
& $({\bf 2}, -\frac12)$
& $({\bf 1}, -1)$ 
& $({\bf 2}, -\frac12)$
& $({\bf 2}, -\frac12)$
& $({\bf 1}, 1)$
& $({\bf 1}, 2)$    \\
\hline
Type I/Y & 
\multirow{2}{*}{$Z_2$-charge A}
& $+$ 
& $+$
& $-$
& $+$
& $-$
& $+$ 
\\
Type II/X 
&
& $+$ 
& $-$
& $-$
& $+$
& $-$
& $+$ 
\\ 
\hline
Type I/Y & 
\multirow{2}{*}{$Z_2$-charge B}
& $+$ 
& $+$
& $-$
& $+$
& $+$
& $+$ 
\\ 
Type II/X & 
& $+$ 
& $-$
& $-$
& $+$
& $+$
& $+$ 
\\ 
\hline
\end{tabular}
\label{Table1}
\end{table}

\begin{figure}[t]
\center
\includegraphics[width=8cm]{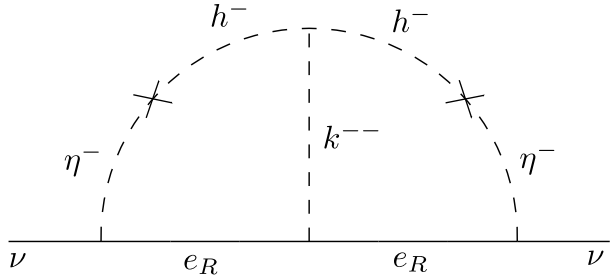}
\caption{
Diagram to induce the neutrino mass in the model with $Z_2$-charge A (Liu-Gu model).
}
\label{fig2}
\end{figure}

\begin{figure}[t]
\center
\includegraphics[width=8cm]{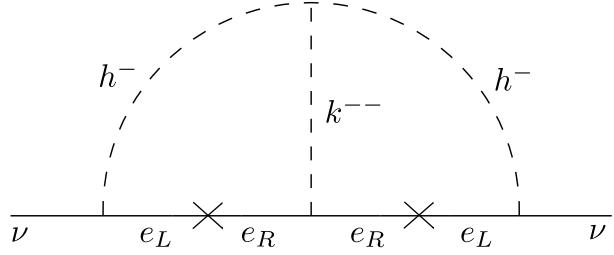}
\caption{
Diagram to induce the neutrino mass in the model with $Z_2$-charge B (Zee-Babu model).
}
\label{fig3}
\end{figure}

We choose the $Z_2$-charge of $k^{++}$ to be $+$ in order to allow the $g$ coupling.
For the $Z_2$-charge assignment of $h^+$, we consider two choices as given in Table \ref{Table1}.
They correspond to the following models:
\begin{enumerate}
\item 
$Z_2$-charge A: Liu-Gu model \cite{Liu:2016mpf}.

The $\ell \ell h^+$ coupling is forbidden.
The $\Phi_1 \Phi_2\, h^+ = \Phi \eta h^+$ coupling is allowed.
The cocktail diagram in Fig.\ref{fig1} induces the $0\nu 2\beta$ decay whose amplitude can be sizable to observe
the process in the near future.
Neutrino masses can be generated by the diagram in Fig.\ref{fig2}.
We note that $h^-k^{++}\Phi_1\Phi_2$ coupling is allowed and
the $0\nu 2\beta$ amplitude via the cocktail diagram 
and the neutrino masses can be also generated via the $h^- k^{++} \Phi_1 \Phi_2$ coupling
in addition to the $h^+h^+k^{--}$ coupling.
The loop-induced neutrino mass matrix is
\begin{equation}
M_\nu \propto M_e g M_e.
\label{MgM}
\end{equation}
The Mu-to-$\mubar$ transition rate is restricted due to the LFV constraints,
as we will explain later.

\item
$Z_2$-charge B: Zee-Babu model \cite{Zee:1980ai,Zee:1985id,Babu:1988ki,Babu:2002uu}.

The $\ell \ell h^+$ coupling is allowed.
The $\Phi_1 \Phi_2 \,h^+$ coupling is forbidden \footnote{We note that only one Higgs doublet is sufficient to radiatively induce the neutrino masses in the original Zee-Babu model.
Even in the model with two Higgs doublets, the structure of this model is essentially the same as the original Zee-Babu model if the $\Phi_1 \Phi_2 h^+$ coupling is absent and the singly charged scalars are not mixed.
Therefore, we simply call this choice the Zee-Babu model.
We believe that this will not cause any confusion to readers.}.
The cocktail diagram in Fig.\ref{fig1} is absent and 
the $0\nu 2\beta$ decay is induced only by the usual contribution from $m_{ee}$.
Neutrino masses can be generated by the diagram in Fig.\ref{fig3},
and
\begin{equation}
M_\nu \propto f M_e g^* M_e f^T.
\label{fMgMf}
\end{equation}
The Mu-to-$\mubar$ transition rate can be sizable to observe in the near-future experiments.

\end{enumerate}

It is important that the $0\nu 2\beta$ decay and Mu-to-$\mubar$ transition
are available to select the models with the $Z$-charges A and B
since only one of their amplitudes can be sizable.
The $\Phi_1 \Phi_2 h^+$ scalar trilinear coupling has a dimension,
and thus, the trilinear coupling can be a soft-breaking term of $Z_2$-charge B.
If the soft-breaking term is added in the model of $Z_2$-charge B, 
it is possible that both amplitudes of the $0\nu 2\beta$ decay and Mu-to-$\mubar$ transition 
are large.
The purpose of this paper is to investigate the possibility
that both can be large enough to be observed in the near future.

\section{Neutrinoless double beta decay via the cocktail diagram}
\label{sec3}

In this section, we describe the $0\nu 2\beta$ contribution from the
cocktail diagram \cite{Gustafsson:2014vpa} in Fig.\ref{fig1}, and
we show the numerical calculation.

In order to execute the triangle loop calculation,
we remark that
the part of the propagator for the neutral scalars in the mass eigenstates can be written as
\begin{equation}
\frac{c^2_{\beta-\alpha}}{m_h^2 - k^2} + \frac{s^2_{\beta-\alpha}}{m_H^2 - k^2}
-\frac{1}{m_A^2 - k^2}
= 
c^2_{\beta-\alpha} \left( \frac{1}{m_h^2 - k^2} - \frac{1}{m_H^2 - k^2}  \right)
+ \frac{1}{m_H^2 - k^2} - \frac{1}{m_A^2 - k^2}.
\end{equation}
If $\cos({\beta-\alpha}) = 0$ and $m_H = m_A$, the contribution to the $0\nu 2\beta$ decay amplitude vanishes.
Since we will work in the alignment limit $\cos({\beta-\alpha}) \to 0$ to avoid experimental constraints, 
we ignore the $h^0/H^0$ part and
only consider the $H^0/A^0$ part in the following.

The operator induced by Fig.\ref{fig1} is
\begin{equation}
{\cal L}_{\rm cocktail} = \frac{A_L}2 (\overline{u_L} \gamma_\mu d_L) (\overline{u_L} \gamma^\mu d_L) (\overline{e_R} (e_R)^c)+{\rm H.c.},
\label{AL-operator}
\end{equation}
and the coefficient is
\begin{align}
A_L &= 8G_F^2 
\frac{1}{16\pi^2}  \left[ c_\phi^2 s_\phi^2
 \left(F_t(1,b,c) + F_t\left(1,\frac{b}{a},\frac{c}{a}\right) - 2 F_t(a,b,c)\right) \frac{m_{hhk}}{m_k^2} g_{ee} \right. 
 \\
& \left. \
- s_\phi c_\phi \left(c_\phi^2 \left(F_t(1,b,c)- F_t(a,b,c)\right) + s_\phi^2 \left(F_t(a,b,c) - F_t\left( 1, \frac{b}{a}, \frac{c}{a} \right) \right) \right) \frac{\lambda_{hk\Phi\Phi} v}{m_k^2} g_{ee}
\right], \nonumber
\end{align}
where
the loop function $F_t$ is given in Appendix \ref{appendix:B}, and
\begin{equation}
a = \frac{m_{h_2^+}^2}{m_{h_1^+}^2}, \quad b = \frac{m_H^2}{m_{h_1^+}^2}, \quad c = \frac{m_A^2}{m_{h_1^+}^2}.
\end{equation}
The current bounds of the $0\nu 2\beta$ half-value periods 
by GERDA \cite{Agostini:2020xta}, CUORE \cite{CUORE:2021mvw}, and Kamland-Zen \cite{KamLAND-Zen:2016pfg} are
\begin{align}
 T_{1/2}^{0\nu} (^{76}{\rm Ge}) &> 1.8 \times 10^{26} \ {\rm years}, \\
 T_{1/2}^{0\nu} (^{130}{\rm Te}) &> 2.2 \times 10^{25} \ {\rm years}, \\
 T_{1/2}^{0\nu} (^{136}{\rm Xe}) &> 1.07 \times 10^{26} \ {\rm years}.
\end{align}
We use numerical values of nuclear matrix elements (NME) given in Ref.\cite{Deppisch:2012nb}.
The half-value period for $^{136}$Xe gives a little stronger bound to the amplitude than those of $^{76}$Ge and
$^{130}$Te due to the NME,
and we obtain the numerical bound in the case of $\lambda_{hk\Phi\Phi} = 0$ as follows:
\begin{equation}
\left|c_\phi^2 s_\phi^2
 \left(F_t(1,b,c) + F_t\left(1,\frac{b}{a},\frac{c}{a}\right) - 2 F_t(a,b,c)\right) 
 g_{ee} \right| \alt
 0.7 \times 10^{-3} \times \frac{{m_k^2}/{m_{hhk}}}{1 \ {\rm TeV}}.
\end{equation}
The experimental constraints for the doubly charged scalar mass are found in 
\cite{Aaboud:2017qph,Dev:2018sel}.
In the following, we suppose $m_k \sim 1$ TeV, which can be allowed by the experiment.

\begin{figure}[t]
\center
\includegraphics[width=12cm]{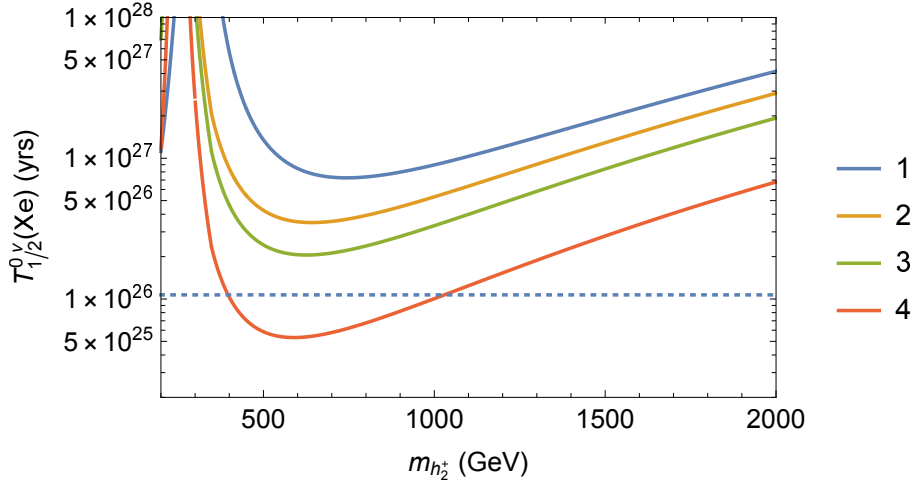}
\caption{
Plot of the $0\nu 2\beta$ half-lives of Xe for various scalar masses.
1. $m_{h_1^+} =300$ GeV, $m_{H} = 450$ GeV, $m_A = 415$ GeV,
2. $m_{h_1^+} =250$ GeV, $m_{H} = 450$ GeV, $m_A = 415$ GeV,
3. $m_{h_1^+} =250$ GeV, $m_{H} = 400$ GeV, $m_A = 360$ GeV,
4. $m_{h_1^+} =250$ GeV, $m_{H} = 300$ GeV, $m_A = 240$ GeV.
Horizontal axis is for the mass of $h_2^+$.
The horizontal-dotted line gives the current bound of the half-life by Kamland-Zen.
}
\label{0nu2b-plot}
\end{figure}

In Fig.\ref{0nu2b-plot}, we plot
the half-value period for Xe as a function of $m_{h_2}^+$
for $m_k^2/m_{hhk} = 1$ TeV, $g_{ee} = 0.3$,
and various masses of the scalars, $m_H$, $m_A$, and $m_{h_1^+}$.
We choose $m_A$ so that $m_H^2 - m_A^2 \simeq v^2$.
We suppose $\lambda_{hk\Phi\Phi}=0$.
As described in Appendix \ref{appendix:A}, the scalar trilinear coupling should be roughly smaller than
the quadratic masses to avoid charge symmetry breaking, and we assume the $\eta^+$-$h^+$ mixing to plot the figure
as
\begin{equation}
\sin2\phi = \frac{2 v m_{\Phi\Phi h}}{m_{h_2^+}^2 - m_{h_1^+}^2}  = {\rm min} \left(\frac{2v}{ m_{h_2^+}},1 \right).
\end{equation}
We note that the amplitude vanishes when $m_{h_1^+} = m_{h_2^+}$.
It is expected that the $0\nu 2\beta$ decay can be observed in the future
if $|g_{ee}| \sim O(0.1)$ and the masses of $H$, $A$, and $h_1^+$ (those come from the doublets, roughly)
are less than several hundreds of GeV.

We note that 
a $0\nu 2\beta$ operator, $(\overline{u_L} d_R)^2 (\overline{e_R}e_R^c)$,
can be generated
at the tree level via a $\overline{u_L} d_R \eta^+$
coupling in the Liu-Gu model.
However, the $\overline{u_L} d_R \eta^+$ coupling is related to the down quark Yukawa coupling,
and the tree-level contribution to the $0\nu 2\beta$ amplitude is less than the loop contribution.
In general 2HDM (so-called the type III model), the $\overline{u_L} d_R \eta^+$ coupling
can be sizable, and the tree-level contribution can be made large compared to the 2HDM with the $Z_2$ charge,
though we do not touch it in this paper.

\section{Loop-induced neutrino mass matrices}
\label{sec4}

The neutrino masses are induced by the diagrams given in Fig.\ref{fig2} and Fig.\ref{fig3}
in the Liu-Gu and Zee-Babu models, respectively,
and the flavor part of the induced neutrino mass matrices are given in
Eqs.(\ref{MgM}) and (\ref{fMgMf}).
In this section, we show more explicit expressions of the neutrino mass matrices
and give the solutions in terms of the $f$ and $g$ coupling matrices to reproduce the observed neutrino oscillations.

The $\eta^-$ couplings to the leptons are given as
\begin{equation}
-{\cal L} \supset r \frac{(M_e)_{ij}}{v}  \overline{e_{iR}} \nu_{jL} \eta^-,
\end{equation}
where
\begin{equation}
r = \left\{ 
 \begin{array}{rl}
  -\cot \beta & (\mbox{for type I/Y}), \\
  \tan\beta & (\mbox{for type II/X}).
 \end{array}
\right.
\end{equation}
In the Liu-Gu model, the loop-induced neutrino mass matrix by the diagram in Fig.\ref{fig2} is
\begin{align}
M_\nu^{\rm LG} &= \frac{4}{v^2}\, r^2  M_e g M_e  \left(s_\phi^2 c_\phi^2 (L_1 (x,x) + L_1(y,y) - 2 L_1(x,y)) m_{hhk} \right. \nonumber   
 \\
&\qquad \qquad \qquad \quad \left. -s_\phi c_\phi (c_\phi^2 (L_1 (x,x) -  L_1(x,y)) + s_\phi^2 (L_1(x,y)- L_1(y,y))) \lambda_{hk\Phi\Phi} v\right),
\end{align}
where
\begin{equation}
x = \frac{m_{h_1^+}^2}{m_k^2},
\qquad
y = \frac{m_{h_2^+}^2}{m_k^2}.
\end{equation}
In the Zee-Babu model without $h^+$\dash$\eta^+$ mixing, the neutrino mass matrix induced by the diagram in 
Fig.\ref{fig3}
is
\begin{equation}
M_\nu^{\rm ZB} = 16 f M_e g^* M_e f^T L_2 (y,y) \frac{m_{hhk}}{m_k^2}.
\end{equation}
The loop functions $L_1$ and $L_2$ are given in Appendix \ref{appendix:C}. 
It can be seen that the flavor-dependent parts are given as in 
Eqs.(\ref{MgM}) and (\ref{fMgMf}),
and the model parameters determine the proportional coefficients.
We note that
a three-loop diagram with a cocktail diagram \cite{Gustafsson:2012vj}
can also induce the neutrino mass
in the Liu-Gu model.
The flavor structure is also given in Eq.(\ref{MgM}).

In the Liu-Gu model, the solution of the $g$ coupling is simply
\begin{equation}
g_{\rm LG} \propto M_e^{-1} M_\nu M_e^{-1}.
\label{LG-sol}
\end{equation}
In the Zee-Babu model, we need a little explanation to show its solution for $g$ and $f$ matrices.
First of all, the $f$ coupling is an antisymmetric matrix, and thus, the rank of the neutrino mass matrix is two,
and the mass matrix (in the normal mass hierarchy) is given as
\begin{equation}
M_\nu = U^* {\rm diag} (0, m_2, m_3) U^\dagger,
\end{equation}
where $U$ is the Pontecorvo-Maki-Nakagawa-Sakata (PMNS) neutrino mixing unitary matrix,
and $m_2$, $m_3$ are neutrino masses (with a Majorana phase in a convention, ${\rm arg}(m_2/m_3) = \alpha_2$).
We write the mixing matrix as
\begin{equation}
U = ( u_1, u_2, u_3),
\end{equation}
where $u_i$'s are column vectors.
We set
\begin{equation}
f_{ij} = f_0 \epsilon_{ijk} (u_1)_k.
\label{fij}
\end{equation}
From the orthogonality of the column vectors, one obtains
\begin{equation}
f U = f_0 ( 0 , -u_3^*, u_2^*) = f_0 U^* 
\left(
\begin{array}{ccc}
 0 & 0 & 0 \\
 0 & 0 & -1 \\
 0 & 1 & 0
\end{array}
\right).
\end{equation}
We find that 
\begin{equation}
g^*_{\rm ZB}  \propto \frac{1}{f_0^2}M_e^{-1} U 
\left(
 \begin{array}{ccc}
  a_1 & a_2 & a_3 \\
  a_2 & m_3 & 0 \\
  a_3 & 0 & m_2
 \end{array}
\right)
U^T M_e^{-1}
\label{ZB-sol}
\end{equation}
gives the algebraic solution of $f M_e g^*_{\rm ZB} M_e f^T \propto M_\nu$ \cite{Fukuyama:2021iyw}.
Due to $f u_1 = 0$, $a_i$'s are free.
The flavor structure of the $g$ coupling in the Zee-Babu model can be controlled by the three free parameters. 
The solution in the inverted mass hierarchy ($m_3=0$) is obtained 
similarly by choosing $f_{ij} = f_0 \epsilon_{ijk} (u_3)_k$.
We consider only the case of the normal mass hierarchy in this paper.

It is clear from this algebraic solution 
that
 all the neutrino oscillation parameters,
 three mixings, a Dirac phase $\delta$, two masses ($m_1 = 0$), and one Majorana phase ($\alpha_2$), can be realized in the Zee-Babu model in principle.
The (1,1) element of the neutrino mass matrix, which is an important parameter for the usual light neutrino exchange diagram for the $0\nu 2\beta$ decay,
is thus simply calculated in the Zee-Babu model as 
\begin{align}
|m_{ee}| &= |m_2 (U_{e2}^*)^2 + m_3 (U_{e3}^*)^2| 
\nonumber
\\
&= 
\left|\sqrt{\Delta m^2_{\rm sol}} \sin^2\theta_{12} \cos^2\theta_{13} + \sqrt{\Delta m^2_{\rm atm} +\Delta m^2_{\rm sol}} e^{i (2\delta-\alpha_2)} \sin^2\theta_{13}\right|,
\end{align}
and we obtain
\begin{equation}
|m_{ee}| \simeq 1 \mbox{\dash}4  \ {\rm meV},
\label{mee-ZB}
\end{equation}
by using the current experimental measurement 
of the oscillation parameters \cite{Esteban:2020cvm}.

\section{Mu-to-$\mubar$ transition and LFV constraints to realize the neutrino matrices}
\label{sec5}

Either the $f$ coupling or the off-diagonal elements of the $g$ coupling matrix 
can induce the LFV decay processes.
In this section, we review the LFV constraints in the Liu-Gu and Zee-Babu models.
Our concern is
the Mu-to-$\mubar$ transition from the exchange of the doubly charged scalar $k^{++}$.
We describe whether the transition rate can be as large as the current experimental bound,
avoiding the LFV constraints. 
There are five independent four-fermion Mu-to-$\mubar$ transition operators \cite{Conlin:2020veq}.
The induced operator by the $k^{++}$ exchange is
\begin{equation}
-{\cal L} = \frac{4G_2}{\sqrt2} (\overline{\mu_R} \gamma_\alpha e_R) (\overline{\mu_R} \gamma^\alpha e_R) +{\rm H.c.},
\label{Lag:G2}
\end{equation}
and 
\begin{equation}
G_2 = - \frac{g_{ee} g_{\mu\mu}^*}{4\sqrt2 \, m_k^2}.
\end{equation}
The bound from the transition experiment \cite{Willmann:1998gd} is
\begin{equation}
\frac{|G_2|}{G_F} \leq 3 \times 10^{-3}.
\end{equation}
The LFV constraints are a bottleneck to obtaining a sizable Mu-to-$\mubar$ transition rate
if the $g$ coupling matrix has off-diagonal elements.
Appendix \ref{appendix:D} summarizes the LFV constraints.
To satisfy the stringent constraint from the $\mu \to 3e$ decay, $g_{e\mu}$ has to be tiny.
The constraint from $\mu \to e\gamma$ decay gives a constraint 
to the size of not only $g_{e\mu}$ but also $g_{e\tau}$ (if $g_{\mu\tau} \neq 0$).

In the case of the Liu-Gu model in Eq.(\ref{LG-sol}),
the (1,3) element of the neutrino mass matrix $M_\nu$
has to be non-zero
to reproduce the 12 and 13 neutrino mixings
since $g_{e\mu}$ is tiny.
Then, $g_{e\tau}$ needs to be sizable, and
consequently,  
$|G_2|/G_F$ is bounded to be less than $O(10^{-5})$ 
due to $\tau \to 3e$ in the Liu-Gu model
(see Eq.(\ref{tau-to-3e}) and Ref.\cite{Fukuyama:2021iyw}).

In the Zee-Babu model, on the other hand,
both $g_{e\mu}$ and $g_{e\tau}$ can be made to be zero
using the free parameters $a_i$ in Eq.(\ref{ZB-sol}).
As a result, $|G_2|/G_F \sim O(10^{-3})$ is possible.
The description of the LFV constraints in the Zee-Babu model is found 
in the literature \cite{Nebot:2007bc,Schmidt:2014zoa,Herrero-Garcia:2014hfa,Alcaide:2017dcx,Irie:2021obo}.
The algebraic solution of the Zee-Babu model is given in Eqs.(\ref{fij}) and (\ref{ZB-sol}).
Since the elements of the $f$ coupling are determined by the first column of the PMNS unitarity matrix,
the factor $f_0$ of the $f$ coupling matrix needs to be small 
due to the $\mu \to e\gamma$ constraint given in Eq.(\ref{fetau-fmutau}).
Then, the elements of the $g$ coupling matrix need to be sizable 
to reproduce the proper neutrino masses for a scalar mass spectrum $m_{h_1^+}, m_k \sim 1$ TeV.
Due to $m_e \ll m_\mu \ll m_\tau$, the $g_{ee}$ element is easily more than unity.
To avoid $g_{ee} \gg 1$, we need to use one of the degrees of freedom of the parameters $a_i$.
Since we need to make $g_{e\mu} \to 0$,
one of $g_{e\tau}$ and $g_{\mu\tau}$ cannot be adjusted.
The restriction from the $\tau^+ \to e^- \mu^+\mu^+$ decay to 
$g_{e\tau}$ element is more stringent due to the charged lepton mass hierarchy.
Consequently, $g_{\mu\tau}$ cannot be made small,
and 
the $\mu \to e\gamma$ decay from the $f$ coupling and the $\tau\to 3\mu$ decay
from the $g$ coupling restrict the mass parameters in the Zee-Babu model.
The Mu-to-$\mubar$ transition rate is bounded by
the $\tau \to 3\mu$ and $\tau^+ \to \mu^- e^+ e^+$ decays,
and the bound of the transition rate is just below the current experimental bound \cite{Fukuyama:2021iyw}.

\section{Analyses for our model}
\label{sec6}

As we have explained,
the cocktail diagram can generate the $0\nu 2\beta$ amplitude
and the $0\nu 2\beta$ half-value periods of Ge and Xe can be just above the current bound
in the Liu-Gu model.
In the Zee-Babu model, on the other hand, 
the $0\nu 2\beta$ amplitude from the cocktail diagram 
is not generated 
and the half-value period is $1000$\dash$10000$ times larger than the current bound
since $\Phi \eta h^+$ coupling
is absent and the $0\nu 2\beta$ decay is induced only by $m_{ee}$ given in Eq.(\ref{mee-ZB}).
The Mu-to-$\mubar$ transition rate is far below the current bound
in the Liu-Gu model due to the LFV constraints, while
it can be just below the current bound in the Zee-Babu model.
These are useful to distinguish the models.

Even in the model with $Z_2$-charge B in Table \ref{Table1} for the Zee-Babu model,
the $\Phi_1 \Phi_2 h^+$ term can be employed as a soft-breaking term in principle.
Then, the cocktail diagram can generate the $0\nu 2\beta$ amplitude,
and the half-value period can be just above the current bound if $g_{ee}$ is sizable.
At that time, the Mu-to-$\mubar$ transition rate can be as large as the current bound
satisfying the LFV constraints,
which we will verify in this section.
\footnote{We missed a non-negligible one-loop diagram in the discussion here.
See the note added for the corrected discussion.}

The two-loop induced neutrino mass matrix in the $Z_2$-charge B model with the soft-breaking $\Phi_1 \Phi_2 h^+$ term
is 
\begin{equation}
M_\nu = r^2 s_{2\phi}^2 \frac{M_e g M_e}{v^2}  m_{hhk} L_{\rm LG}
+ 16 f M_e g^* M_e f^T \frac{m_{hhk}}{m_k^2} L_{\rm ZB} ,
\label{eq:neutrino_mass_two-loop}
\end{equation}
where $L_{\rm LG}$ and $L_{\rm ZB}$ are given in Appendix \ref{appendix:C}.
The dimensionless coupling $\lambda_{hk \Phi \Phi}$ is forbidden by the $Z_2$ charge.
This equation is rewritten as
\begin{equation}
R M_e g  M_e + f M_e g^* M_e f^T = \frac{M_\nu}{16 L_{\rm ZB}}  \frac{m_k^2}{m_{hhk}},
\label{R-equation}
\end{equation}
where
\begin{equation}
R = \frac{r^2 s_{2\phi}^2 m_k^2}{16v^2} \frac{L_{\rm LG}}{L_{\rm ZB}}.
\end{equation}
We comment that there can be a three-loop contribution in the Liu-Gu model,
and the parameter $R$ can be redefined by containing the contribution in Eq.(\ref{R-equation}).

We assume $g_{e\mu}= g_{e\tau} =0$ to satisfy the LFV bound and fix the value of $g_{ee}$ to give a contribution
to the $0\nu 2\beta$ decay by the cocktail diagram.
Then, there are three parameters in $g$ and three parameters in $f$ coupling matrices.
Equation~(\ref{R-equation}) contains six equations.
Therefore, for fixed values of $R$, $L_{\rm ZB}$, $m_k^2/m_{hhk}$, and neutrino oscillation parameters in $M_\nu$,
one can solve the equations for $g_{\mu\mu}$, $g_{\mu\tau}$, $g_{\tau\tau}$,
$f_{e\mu}$, $f_{e\tau}$, and $f_{\mu\tau}$.
The solutions are screened by the LFV constraints,
especially for $|f_{\mu\tau} f_{e\tau}| $ from the $\mu \to e\gamma$ bound.
From the assumption that $g_{e\mu} = g_{e\tau} = 0$, 
the solar neutrino mixing and 13-mixing have to be generated from $f$.
Therefore, roughly saying, $R$ should be smaller than $O(0.1)$ to satisfy the $\mu \to e\gamma$ bound.

To consider the LFV constraints of the model parameters, the size of $g_{\mu\mu}$ is important.
Roughly speaking, the numerical size of $g_{\mu\mu}$ is determined by the (2,2) element of Eq.(\ref{R-equation}) as
\begin{equation}
R g_{\mu\mu}  \sim (M_\nu)_{22} \frac{1}{16 m_\mu^2 L_{\rm ZB}} \frac{m_k^2}{m_{hkk}},
\end{equation}
where we ignore $O(f_{12}^2), O(f_{13}^2)$ terms, which are small in the solution.
The (2,2) element of the neutrino mass is roughly fixed by the neutrino oscillation data as 
\begin{equation}
(M_\nu)_{22} \sim \sin^2\theta_{\rm atm} \times 0.05 \ {\rm eV}.
\end{equation}
To reproduce the atmospheric neutrino mixing, 
one needs
\begin{equation}
g_{\mu \tau} \sim \frac{m_\mu}{m_\tau} g_{\mu\mu}.
\end{equation}
As a result, the $\tau^+ \to \mu^- e^+ e^+$ decay process constrains 
the solution of a smaller $R$ for fixed $m_k$.
The size of $G_2/G_F$ is constrained to be less than $O(10^{-3})$
from the $\tau^+ \to \mu^- e^+ e^+$ and $\tau \to 3\mu$ decay processes.
See Eqs.(\ref{tau-to-muee}) and (\ref{tau-to-3mu}).

\begin{figure}[t]
\center
\includegraphics[width=10cm]{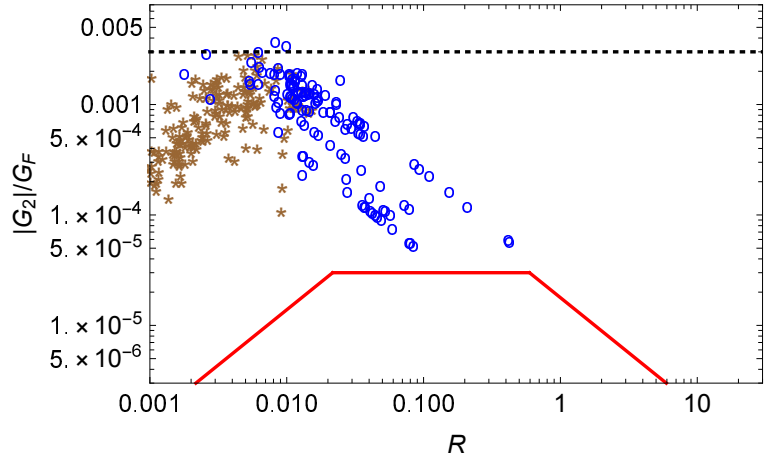}
\caption{
$|G_2|/G_F$ as a function of $R$.
The solid line shows the bound in the case where the Zee-Babu contribution is zero (the $\ell \ell h$ coupling is zero).
For $R < O(1)$, the LFV decay constrains $|G_2|/G_F$.
Even though $g_{\mu\mu}$ becomes larger for smaller $R$, $|G_2|/G_F$ cannot become larger.
The circles ``o" are plotted 
if the corresponding parameters to realize the neutrino mass matrix can satisfy the LFV constraints
by switching on the $\ell \ell h$ coupling.
The asterisks ``\textasteriskcentered'' are plotted if 
the doubly charged scalar $k^{++}$ needs to be heavier than 1 TeV to satisfy the LFV data. 
The horizontal-dotted line shows the current experimental bound from the Mu-to-$\mubar$ transition.
See the text for more detail on the setup.
}
\label{fig:G2bound}
\end{figure}

In Fig.\ref{fig:G2bound}, we show a plot of $G_2/G_F$ as a function of $R$
to depict the statements above.
The solid line shows the plot 
in the case of the Liu-Gu model ($f=0$ in Eq.(\ref{R-equation})).
The (1,2) element of $M_\nu$ is chosen to be zero 
(namely $g_{e\mu} = 0$ to satisfy the stringent $\mu\to 3e$ constraint),
which means that the neutrino oscillation parameters are related (see Ref.\cite{Fukuyama:2021iyw}).
The ratio of $g_{e\tau}/g_{\mu\mu}$ is fixed as a solution to reproduce $M_\nu$.
We choose $g_{ee} = 0.3$ and $m_k = 1$ TeV for $R\agt O(1)$.
For $R \alt O(1)$, it violates
the bound of $\tau \to 3e$ process,
and therefore, we need to enlarge $m_k$ to satisfy the bound.
Then, the solid line becomes flat for $R \alt O(1)$,
which can be understood by Eq.(\ref{tau-to-3e}) for fixed $g_{e\tau}/g_{\mu\mu}$.
We note that the $\mu \to e\gamma$ decay can restrict $G_2/G_F$
by Eq.(\ref{mueg-doubly-charge}) for smaller values of $R$.
The circles are plotted as the solutions of 
Eq.(\ref{R-equation}) that satisfies the LFV constraints 
for
$g_{ee} = 0.3$, $m_{h_2^+} = 1$ TeV, $L_{\rm ZB} = 2/(16\pi^2)^2$, and $m_{hhk} = m_k = 1$ TeV.
In order to obtain the points, 
the neutrino mass and mixing parameters \cite{Esteban:2020cvm}, as well as phases, are randomly generated 
within experimental errors.
The asterisks are plotted as the solutions
similarly for 
$g_{ee} = 0.3$, $m_{h_2^+} = 1$ TeV, $L_{\rm ZB} = 2/(16\pi^2)^2$, and $m_k^2/m_{hhk} = 1$ TeV.
The $\tau^+ \to \mu^- e^+ e^+$ and $\tau \to 3\mu$ bounds 
are satisfied by choosing $m_k$ ($> 1$ TeV).
In those plotted solutions, the $0\nu 2\beta$ contribution from the cocktail diagram can
be a little above the current experimental bound if the scalars in the doublets are light as shown in Section \ref{sec3}.

In order to make clear the plots of the solutions of Eq.(\ref{R-equation}) in Fig.\ref{fig:G2bound},
we exhibit the numerical examples of the solutions
to reproduce the best fit values of the neutrino mixing angles and mass squared differences 
by NuFIT 5.1 \cite{Esteban:2020cvm}.
\begin{enumerate}
\item  $R = 0.01$:
\begin{align}
g &= \left(
 \begin{array}{ccc}
  0.3 & 0 & 0 \\
  0 & 0.36903 &  0.0061989 \\
  0 & 0.0061989 & 0.0011524 - 0.0006303 i
 \end{array}
\right), \\
f &= 
\left(
 \begin{array}{ccc}
  0 & 0.041029 - 0.015809 i & 0.0010437 + 0.0088745 i \\
  -0.041029 + 0.015809 i & 0 & 0.022215 + 0.060334 i\\
  -0.0010437 - 0.0088745 i & -0.022215 - 0.060334 i & 0
 \end{array}
\right).
\end{align}
For $m_k = 1$ TeV, the branching ratios of the three-body LFV decays are calculated 
from Eq.(\ref{3-body-LFV}) as 
\begin{equation}
\{{\rm Br} (\tau \to 3\mu), {\rm Br} (\tau^+ \to \mu^- e^+ e^+)\} 
= \{1.7, 1.1\} \times 10^{-9},
\end{equation}
and the experimental bounds given in Eqs.(\ref{bound:tau-decay-1}) and (\ref{bound:tau-decay-2})
are satisfied in this solution.
We obtain $|G_2|/G_F = 1.7 \times 10^{-3}$ for $m_k = 1$ TeV.

\item  $R = 0.005$:
\begin{align}
g &= \left(
 \begin{array}{ccc}
  0.3 & 0 & 0 \\
  0 & 0.32538 &  0.035573 \\
  0 & 0.035573 & -0.0004850 - 0.0012792 i
 \end{array}
\right), 
\\
f &= 
\left(
 \begin{array}{ccc}
  0 & 0.005829 + 0.050616 i & -0.021693 + 0.008959 i \\
  -0.005829 - 0.050616 i & 0 & -0.027699 - 0.008293 i \\
 0.021693 - 0.008959 i & 0.027699 + 0.008293 i  & 0
 \end{array}
\right).
\end{align}
For $m_k = 1$ TeV, the branching ratios are calculated as
\begin{equation}
\{{\rm Br} (\tau \to 3\mu), {\rm Br} (\tau^+ \to \mu^- e^+ e^+)\} 
= \{4.3, 3.6\} \times 10^{-8},
\end{equation}
and they exceed the experimental bounds in Eqs.(\ref{bound:tau-decay-1}) and (\ref{bound:tau-decay-2}).
To satisfy the bounds, one needs $m_k \agt 1.25$ TeV.
We obtain $|G_2|/G_F < 0.95 \times 10^{-3}$ for $m_k > 1.25$ TeV.

\end{enumerate}
In those solutions, the branching ratios of the $\mu \to e\gamma$ decay
are smaller than the experimental bound
in our setup of the singly charged scalar mass spectrum
due to $|f_{e\tau} f^*_{\mu\tau}| < 0.001$ (see Appendix \ref{appendix:D}).

In our model of $Z_2$-charge B with $\Phi_1 \Phi_2 h^+$ coupling,
both the Mu-to-$\mubar$ transition rate and the $0\nu 2\beta$ amplitudes
can be large for $R \simeq O(0.001)$\dash$O(0.1)$.
This size of $R$ can be realized if $r< 1$, which corresponds to the type I/Y 2HDM.
In fact, the experimental constraints of the scalar decays to fermions
can allow light scalars in the doublets \cite{Aiko:2020ksl},
especially in the type I 2HDM.
On the other hand, the Liu-Gu model with $Z_2$-charge A needs a larger $r$
to reproduce the neutrino masses while suppressing the LFV decays for $m_k \sim m_{hhk} \sim 1$ TeV,
and then, the light scalars are constrained from the experimental data of the scalar decays to fermions.
The lighter scalars can produce the smaller half-value periods as shown in Fig.\ref{0nu2b-plot}.

\section{Extension to the left-right model}
\label{sec7}

In the left-right model whose gauge symmetry is $SU(3)_c \times SU(2)_L \times SU(2)_R \times U(1)_{B-L}$,
the singly and doubly charged scalars can be unified in the $SU(2)_R$
triplet $\Delta$ (with $B-L = 2$),
\begin{equation}
\Delta^a{}_b = \left(
 \begin{array}{cc}
  \frac1{\sqrt2} \Delta^+ & \Delta^0 \\
  \Delta^{++} & - \frac1{\sqrt2} \Delta^+
 \end{array}
\right).
\end{equation}
The two $SU(2)_L$ doublets are unified into a bidoublet, $({\bf 2}, {\bf 2}, 0)$ under $SU(2)_L \times SU(2)_R \times U(1)_{B-L}$.
A right-handed neutrino and a right-handed charged lepton form a $SU(2)_R$ doublet. 
The $W_R$ gauge boson exchanges can induce new tree-level contributions to the $0\nu 2\beta$ decay \cite{Mohapatra:1980yp,Mohapatra:1981pm,Picciotto:1982qe,Hirsch:1996qw,Tello:2010am,Nemevsek:2011aa,Barry:2012ga,Parida:2012sq,Barry:2013xxa,Dev:2014xea,Aydemir:2014ama,Deppisch:2014zta,Borah:2016iqd,Pritimita:2016fgr,Borah:2016hqn}.
However, recent experimental data push up the bound of the $W_R$ mass \cite{Nemevsek:2018bbt,CMS:2021dzb,Aaboud:2019wfg}.
As a result, the amplitude from the one-loop cocktail diagram can be larger than the amplitudes at the tree-level 
$W_R$ exchanges.
In this section, we discuss those contributions with a new box loop contribution 
in the model of the extension to the left-right model.

\begin{figure}[t]
\center
\includegraphics[width=8cm]{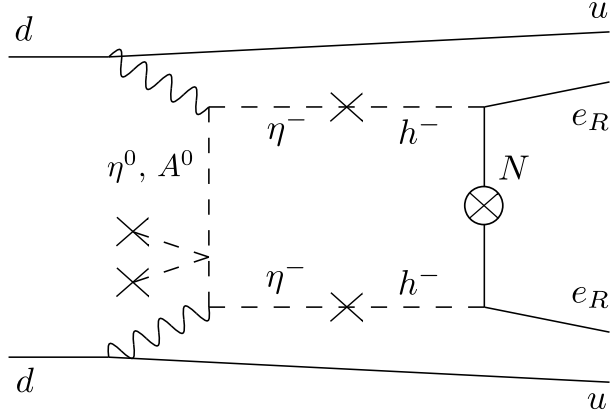}
\caption{
Box loop diagram to induce the $0\nu 2\beta$ decay.
The $\times$ mark stands for the insertion of the electroweak vev,
and the $\otimes$ mark stands for the mass insertion of $N$. 
}
\label{fig5}
\end{figure}

Before going to the left-right model, we simply add a singlet fermion $N$ to the 2HDM with $h^+$ and $k^{++}$.
The coupling and mass of $N$ are given as 
\begin{equation}
\left(\kappa\, \overline{e_R} N h^-+{\rm H.c.}\right)  + \frac12 m_N \overline{N^c} N.
\end{equation}
The $\kappa$ coupling can induce the $0\nu 2\beta$ operator from the box diagram in Fig.\ref{fig5}.
The coefficient of the operator in Eq.(\ref{AL-operator}) is
\begin{equation}
A_L^{\rm Fig.\ref{fig5}} = 8G_F^2
\frac{1}{16\pi^2} c_\phi^2 s_\phi^2
\left(F_b(1,b,c,d) + \frac1{a}F_b\left(1,\frac{b}{a},\frac{c}{a},\frac{d}{a}\right) - 2 F_b(a,b,c,d)\right) \frac{m_N}{2m_{h_1^+}^2} \kappa^2, 
\end{equation}
where
\begin{equation}
a = \frac{m_{h_2^+}^2}{m_{h_1^+}^2}, \quad b = \frac{m_H^2}{m_{h_1^+}^2}, \quad c = \frac{m_A^2}{m_{h_1^+}^2},
\quad
d = \frac{m_N^2}{m_{h_1^+}^2},
\end{equation}
and the loop function $F_b$ is given in Appendix \ref{appendix:B}.
We comment that one can also consider the Dirac neutrino coupling, $\ell N \Phi$.
However, the coupling can generate too large neutrino masses 
(even if it only couples to the inert doublet, $\ell N \eta$, and the Dirac neutrino mass is absent \cite{Ma:2006km}),
and therefore, the coupling should be too small to contribute to the $0\nu 2\beta$ amplitude.

\begin{figure}
\center
\includegraphics[width=10cm]{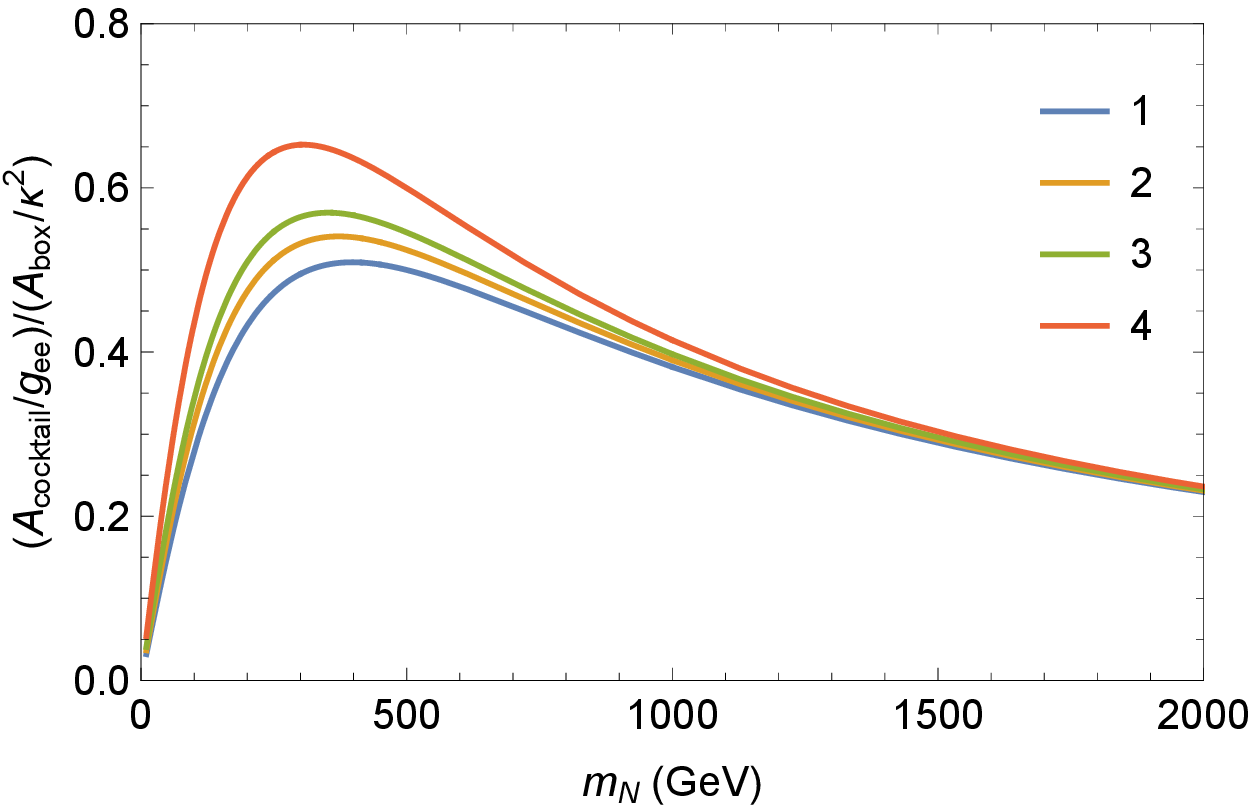}
\caption{
Ratio of the amplitudes of $A_{\rm cocktail}/g_{ee}$ and $A_{\rm box}/\kappa^2$
for various scalar masses.
1. $m_{h_1^+} =300$ GeV, $m_{H} = 450$ GeV, $m_A = 415$ GeV,
2. $m_{h_1^+} =250$ GeV, $m_{H} = 450$ GeV, $m_A = 415$ GeV,
3. $m_{h_1^+} =250$ GeV, $m_{H} = 400$ GeV, $m_A = 360$ GeV,
4. $m_{h_1^+} =250$ GeV, $m_{H} = 300$ GeV, $m_A = 240$ GeV.
We choose $m_k^2/m_{hhk} = 1$ TeV and $m_{h_2^+} = 1$ TeV.
The horizontal axis is the mass of $N$.
}
\label{fig:ratio-cocktail-box}
\end{figure}

In Fig.\ref{fig:ratio-cocktail-box}, we plot the ratio of the amplitudes from the cocktail and box diagrams.
We choose $m_k^2/m_{hhk} = 1$ TeV and $m_{h_2^+} = 1$ TeV in the plot.
The box loop can give a subleading contribution to the $0\nu 2\beta$ amplitude.
If $m_{hhk}/m_k^2 \ll 1/(1 \ {\rm TeV})$,
the cocktail contribution is suppressed and the box contribution can become the leading contribution
to the decay amplitude.

\begin{figure}[t]
\center
\includegraphics[width=4.4cm]{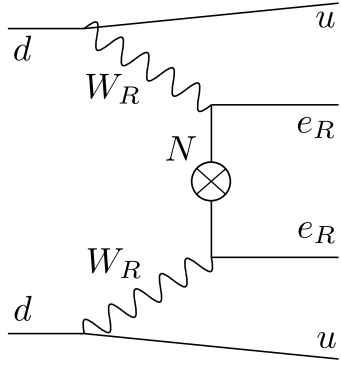}
\caption{
Diagram to induce the $0\nu 2\beta$ decay by the $W_R$ currents with the exchange of the right-handed neutrino $N$.
The $\otimes$ mark stands for the Majorana mass insertion.
}
\label{fig6}
\end{figure}

\begin{figure}[t]
\center
\includegraphics[width=4.4cm]{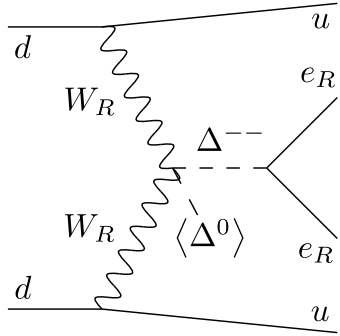}
\caption{
Diagram to induce the $0\nu 2\beta$ decay by the gauge interaction of the $SU(2)_R$ triplet $\Delta$.
}
\label{fig7}
\end{figure}

Now we consider the left-right model.
The right-handed charged lepton and the SM singlet fermion $N$ are contained in a $SU(2)_R$ doublet $\ell_{R}$,
and the $g$ coupling in Eq.(\ref{f-and-g}) is extended as
\begin{equation}
g_{ij} \overline{\ell_{iR}{}^c} \ell_{jR} \Delta
= g_{ij} (\overline{e_{iR}{}^c} e_{jR} \Delta^{++} + \sqrt{2} \, \overline{e_{iR}{}^c} N_j \Delta^+
 - \overline{N^c_i} N_j \Delta^0).
\end{equation}
As we have mentioned at the beginning of this section,
there are tree-level contributions from the $W_R$ gauge boson to the following $0\nu 2\beta$ operator:
\begin{equation}
\frac{A_R}2 (\overline{u_R} \gamma_\mu d_R )(\overline{u_R} \gamma^\mu d_R )(\overline{e_R} e_R^c)+{\rm H.c}.
\end{equation}
Let us estimate the tree-level contributions for the current experimental bound of $M_{W_R} \agt 5$ TeV.
Since our purpose here is 
to compare the tree-level contributions with the one-loop contributions,
we here describe the contributions only from the diagrams in Figs.\ref{fig6} and \ref{fig7}.
Surely, there are additional contributions that pick $W_L$\dash$W_R$ mixing.
The contributions from the diagrams are
\begin{align}
A_R^{\rm Fig.\ref{fig6}} &= 8 G_F^2 \left( \frac{g_R^2}{g_L^2}\frac{M^2_{W_L}}{M^2_{W_R}} \right)^2 \frac{1}{ m_N}, \\
A_R^{\rm Fig.\ref{fig7}} &= 
8 G_F^2 \left( \frac{g_R^2}{g_L^2}\frac{M^2_{W_L}}{M^2_{W_R}} \right)^2 \frac{4\langle \Delta^0 \rangle}{m_{\Delta^{++}}^2} g_{ee}.
\end{align}
We note that the $W_R^-$--$W_R^-$--$\Delta^{++}$ term is given by the gauge interaction 
\begin{equation}
{\cal L} \supset 
\left| 
g_R ({\bm{ W}}_R \Delta - \Delta {\bm W}_R )
\right|^2 =
g_R^2 W_R^- W_R^- \Delta^{++} \Delta^0 + \cdots,
\end{equation}
where ${\bm W}_R = \frac{\tau^a}2 W^a_R$ and $\tau^a$'s are the Pauli matrices.
We also note that $m_N = 2 g_{ee} \langle \Delta^0 \rangle$
if additional singlet fermions are not introduced and the heavy neutrino mixings are absent.
We obtain
\begin{align}
\frac{A_R^{\rm Fig.\ref{fig6}}}{A_R^0}
&= 0.15 \times \frac{g_R^4}{g_L^4}\left(\frac{5\ {\rm TeV}}{M_{W_R}}\right)^4  \frac{100 \ {\rm GeV}}{m_N}, \\
\frac{A_R^{\rm Fig.\ref{fig7}}}{A_R^0}
&= 0.15 \times \frac{g_R^4}{g_L^4}\left(\frac{5\ {\rm TeV}}{M_{W_R}}\right)^4 \frac{\langle \Delta^0 \rangle}{8 \ {\rm TeV}}
\left(\frac{1\ {\rm TeV}}{m_{\Delta^{++}}}\right)^2 \frac{g_{ee}}{0.3}, 
\end{align}
where $A_R^0$ is the coefficient of the operator that gives the current experimental bound of the $0\nu 2\beta$ 
half-value period of Xe
by using the NME for the operator given in Ref.\cite{Deppisch:2012nb}.
We note that $M_{W_R} \simeq g_R \langle\Delta^0\rangle$ if the vev of $\Delta^0$ dominantly breaks $SU(2)_R \times U(1)_{B-L}$.

In the minimal model where the vev of $\Delta^0$ is the only source to break $SU(2)_R \times U(1)_{B-L} \to U(1)_Y$,
$\Delta^+$ is absorbed to the $W_R$ boson.
The $\eta^-$\dash$\Delta^-$ mixing $\phi$ is determined only by the vevs
irrespective of the detail of the scalar potential.
We obtain
\begin{equation}
\tan\phi = -\frac{g_R}{g_L}\frac{M_{W_L}}{M_{W_R}} \cos2\beta.
\end{equation}
See Appendix \ref{appendix:E} for more details.
As a consequence, 
the contribution from the cocktail diagram via $W_L$ boson exchanges
cannot overcome the contribution from the tree-level diagram in Fig.\ref{fig7}.
If one introduces two $SU(2)_R$ triplets that have vevs to break $SU(2)_R \times U(1)_{B-L}$
or an additional $SU(2)_R$ doublet $({\bf 1}, {\bf 2}, 1/2)$ under ($SU(2)_L$, $SU(2)_R$, $(B-L)/2$),
the $\eta^-$\dash$\Delta^-$ mixing depends on the detail of the scalar couplings.
For example, in the inverse seesaw model \cite{Mohapatra:1986aw} 
which is often considered 
to obtain the tiny neutrino masses more naturally in the TeV-scale left-right model,
the additional $SU(2)_R$ doublet is motivated to be introduced.
If the vev of the $SU(2)_R$ doublet mainly generates the $W_R$ boson mass
and the vev of $\Delta^0$ is smaller than it, the $\eta^-$\dash$\Delta^-$ mixing can become 
a free parameter (independent from the gauge boson mass ratio) in the non-minimal model.
Then, 
as in the 2HDM with $h^+$ and $k^{++}$,
the amplitude of the one-loop contribution from the cocktail diagram can be as large as the current bound,
and
can overcome the contribution from the tree-level $W_R$ exchanges.
The box loop (Fig.\ref{fig5}) (with $\kappa = \sqrt2 g_{ee}$) can also overcome the contribution from the right-handed neutrino in Fig.\ref{fig6} 
if $m_N$ is heavier than several hundred GeV.
We note that
there is a $W_L^+$\dash$W_R^-$\dash$\eta^0$ term from the gauge interaction
and the $W_R$ boson can propagate in the loops in the cocktail and box diagrams, 
though those contributions to the $0\nu 2\beta$ decay amplitude are small
due to the heaviness of the $W_R$ boson.
Surely, there are also tree-level diagrams of the heavy Majorana neutrino $N$ 
exchanges with active-sterile neutrino mixings \cite{Hirsch:1996qw,Tello:2010am,Nemevsek:2011aa,Barry:2012ga,Parida:2012sq,Barry:2013xxa,Dev:2014xea,Aydemir:2014ama,Deppisch:2014zta,Borah:2016iqd,Pritimita:2016fgr}.
However, 
the active-sterile mixings need to be enlarged to obtain the sizable amplitudes.
The large active-sterile mixing can induce the active neutrino masses at the loop level \cite{Pilaftsis:1991ug},
and then, the Majorana mass of $N$ should not be large to obtain the sizable $0\nu 2\beta$ amplitude 
in those cases to utilize the large active-sterile mixings \cite{Mitra:2011qr}
and the box loop contribution can be more significant for $m_N \agt O(100)$ GeV.

In the left-right model,
the tiny active neutrino masses can be generated at the tree level
even if the $g$ coupling matrix is diagonal and the LFV constraints are avoided. 
Therefore, the Mu-to-$\mubar$ transition operator via the tree-level $\Delta^{++}$ exchange
can be as large as the current experimental bound if the doubly charged scalar mass is about 1 TeV,
as discussed in Ref.\cite{Fukuyama:2021iyw}.
The Yukawa interactions of the quarks and leptons,
\begin{equation}
q_L : ({\bf 3},{\bf 2},{\bf 1},\frac16), \quad
q_R : ({\bf 3},{\bf 1},{\bf 2},\frac16), \quad
\ell_L : ({\bf 1},{\bf 2},{\bf 1},-\frac12), \quad
\ell_R : ({\bf 1},{\bf 1},{\bf 2},-\frac12), \quad
\end{equation}
with a bidoublet Higgs multiplet $\Phi_b: ({\bf 1},{\bf2},{\bf2},0)$
under $(SU(3)_c, SU(2)_L, SU(2)_R,(B-L)/2)$
are given as
\begin{equation}
 \overline{q_L} q_R (Y_{q1} \Phi_b + Y_{q2} \tilde\Phi_b) +  \overline{\ell_L} \ell_R (Y_{\ell 1} \Phi_b + Y_{\ell 2}\tilde\Phi_b) + {\rm H.c.}
\end{equation}
The up- and down-type quarks' mass matrices are linear combinations of $Y_{q1}$ and $Y_{q2}$,
and the charged leptons and the Dirac neutrino mass matrices are
linear combinations of $Y_{\ell 1}$ and $Y_{\ell 2}$.
In the inverse seesaw model, we introduce three generations of gauge singlet fermions $S_i$
and a $SU(2)_R$ doublet, $H_R : ({\bf 1},{\bf1},{\bf 2},1/2)$.
The interactions,
\begin{equation}
g_{ij} \overline{\ell_{iR}{}^c} \ell_{jR} \Delta + d_{ij} \overline{S_i^c} \ell_{jR} H_R + \frac12 (\mu_S)_{ij} \overline{S^c_i} S_j
+ {\rm H.c.},
\end{equation}
give the active neutrino mass matrix as
\begin{equation}
M_\nu \simeq m_D m_S^{-1} \mu_S (m_S^T)^{-1} m_D^T,
\end{equation}
for small elements of $\mu_S$,
where $m_D$ is a Dirac neutrino mass matrix,
and $m_S = d \langle H_R^0 \rangle$.
As described, for $\langle \Delta^0 \rangle \ll \langle H_R^0 \rangle$,
the $H_R^+$ boson mainly becomes the longitudinal component of $W_R$ boson, and 
the $\Delta^+$ and $\Delta^{++}$ components in the $SU(2)_R$ triplet $\Delta$
respectively behave as $h^+$ and $k^{++}$ to induce the $0\nu 2\beta$ decay via the cocktail diagram
in Fig.\ref{fig1}.

\section{Conclusion}
\label{sec8}

The $0\nu 2\beta$ decay and the Mu-to-$\mubar$ transition are investigated 
in the 2HDM with $SU(2)_L$ singlet scalars,
$h^+$ and $k^{++}$.
The current bounds of the $0\nu 2\beta$ decay translate to
$m_{ee} \alt 0.1$ eV 
if the ordinary diagram of the Majorana neutrino exchange dominates the decay amplitude.
It is well known that
the observation of the decay just above the current bounds of the half-lives
does not necessarily mean $m_{ee} \sim 0.1$ eV in the models beyond the SM.
The $e_R e_R k^{++}$ and $k^{++} h^- h^-$ interactions, as well as the mixing between $h^+$
and a charged scalar in the $SU(2)_L$ Higgs doublet,
can violate the lepton number.
The $0\nu 2 \beta$ decay can be induced via the diagram given in Fig.\ref{fig1},
and the half-life can be as large as the current bound even if $m_{ee}$ is small.
We have shown that
both the $0\nu 2 \beta$ decay and the Mu-to-$\mubar$ transition rate
can be as large as the current experimental bounds
in the general setup of the 2HDM with $h^+$ and $k^{++}$
that induces the neutrino mass matrix by two-loop diagrams in Figs.\ref{fig2} and \ref{fig3},
while the experimental constraints from the LFV decays of the charged leptons
are satisfied.
The search for both the $0\nu 2\beta$ decay and the Mu-to-$\mubar$ transition
will be important to select models and specify the parameter space of the model.

If we consider another possibility that the Mu-to-$\mubar$ transition is induced in connection with neutrinos,
the type-II seesaw model immediately comes to mind.
The doubly charged scalar is contained in the $SU(2)_L$ triplet $\Delta_L$ for the type-II seesaw term,
and $\ell \ell \Delta_L$ terms can induce an operator 
for the Mu-to-$\mubar$ transition.
However, the constraints of the LFV decays require neutrino mass degeneracy
(even in the case of the normal mass hierarchy)
to obtain the sizable Mu-to-$\mubar$ transition operator.
Consequently, the cosmological bound of a total of the neutrino masses \cite{Aghanim:2018eyx} gives a limitation 
on the Mu-to-$\mubar$ transition rate (see Refs.\cite{Han:2021nod,Fukuyama:2021iyw}).
If the type-I seesaw contribution is added, both the $0\nu 2\beta$ decay amplitude
and the Mu-to-$\mubar$ transition rate can be as large as the current bound. 
However, one needs a cancellation in $m_{ee}$ between the type-I and type-II seesaw contributions, 
which impairs the predictivity of the model.
In the model we propose in this paper,
the amplitudes of the $0\nu 2\beta$ decay and the Mu-to-$\mubar$ transition are
 proportional to the $e_R e_R k^{++}$ coupling (without a cancellation),
and the scalar spectrum is more predictable to induce the half-lives of the $0\nu 2\beta$ decay
and the Mu-to-$\mubar$ transition.
The experimental results at the LHC and ILC will give us information on the possible scalar spectrum \cite{Nomura:2017abh,Crivellin:2018ahj,BhupalDev:2018tox,Padhan:2019jlc,Fuks:2019clu,Bai:2021ony,Ashanujjaman:2021txz,Dey:2022whc}.
This paper claims that the Mu-to-$\mubar$ transition, which will be updated in the near-future experiments,
 can be one of the players to sort out those models which induce the $0\nu 2\beta$ decay.

\section*{Note added}

We correct the discussion in Section~\ref{sec6}.
The induced neutrino mass $M_\nu$ should include not only the two-loop contributions described by Eq.~\eqref{eq:neutrino_mass_two-loop}, but also a one-loop one.
Even after the correction, our main claim is unchanged that the measurement of the Mu-to-$\overline{\rm Mu}$ transition is important to sort out the models which induce the $0\nu 2 \beta$ decay.

\begin{figure}[h]
\center
\includegraphics[width=8cm]{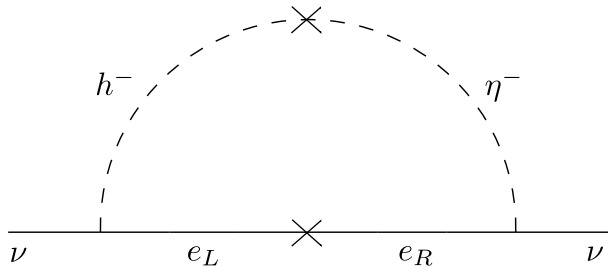}
\caption{
One-loop diagram to induce the neutrino mass in the model with the breaking term of $Z_2$-charge.
}
\label{fig12}
\end{figure}
As discussed in Section~\ref{sec2},
the $f_{ij} \ell_i \ell_j h^+$ coupling is forbidden by the $Z_2$-charge A (given in Table~1),
and only ``Liu-Gu'' contribution in Eq.~\eqref{MgM} is generated by the two-loop diagram in Fig.~\ref{fig2}.
On the other hand, 
the $\Phi_1 \Phi_2 h^+$ scalar trilinear term is forbidden by the $Z_2$-charge B,
and only ``Zee-Babu'' contribution in Eq.~\eqref{fMgMf} is generated by the two-loop diagram in Fig.~\ref{fig3}.
The loop-induced effective neutrino mass operators are
$(\ell \Phi_1^\dagger) (\ell \Phi_1^\dagger)$ or $(\ell \Phi_2^\dagger) (\ell \Phi_2^\dagger)$.
We have introduced a breaking term of the $Z_2$ discrete symmetry
and mix the Liu-Gu and Zee-Babu contributions comparably 
to see if both the Mu-to-$\overline{\rm Mu}$ transitions and $0\nu 2\beta$ decay width can be 
large enough to be observed by near-future experiments.
However, when the breaking term is introduced,
an effective $(\ell \Phi_1^\dagger) (\ell \Phi_2^\dagger)$ operator can be generated
by an one-loop diagram shown in Fig.~\ref{fig12}, as discussed in the original Zee model,
\begin{equation}
M_{\nu}^{\rm Zee}= \frac{{r s_{2\phi}}}{16\pi^2 v} \ln \frac{m_{h_1^+}^2}{m_{h_2^+}^2}
\left(f M_e^2 + M_e^2 f^T \right).
\end{equation}
If the two-loop Liu-Gu and Zee-Babu contributions are comparable, 
the one-loop Zee contribution dominates the neutrino mass matrix.
As it is known, the original Zee model alone cannot reproduce the neutrino oscillation parameters.

The proper treatment for our purpose is, therefore,
that we introduce a small $f_{e\tau}$ coupling to the Liu-Gu model;
the two-loop Liu-Gu and the one-loop Zee contributions are comparable
(the two-loop Zee-Babu contribution is tiny).
Then, the (1,3) element of $M_\nu$ is generated by the one-loop Zee contribution,
and the observed neutrino oscillation parameters can be reproduced 
even in the case of $g_{e\mu} = g_{e\tau}=0$ to avoid too large LFV decay rates.
Therefore, the Mu-to-$\overline{\rm Mu}$ transition can become large, as discussed in the text.
The situation is much simpler to solve the equation given in Eq.~\eqref{R-equation}.

Thus, the third and subsequent paragraphs in Section~\ref{sec6} should be replaced with the following sentences:
The induced neutrino mass in the $Z_2$-charge B model with the soft-breaking $\Phi_1\Phi_2h^+$ term is
\begin{align}
M_\nu=\frac{{r s_{2\phi}}}{16\pi^2 v} \ln \frac{m_{h_1^+}^2}{m_{h_2^+}^2}\left(f M_e^2 + M_e^2 f^T \right)+r^2s_{2\phi}^2\frac{M_egM_e}{v^2}m_{hhk}L_\mathrm{LG},
\label{eq:neutrino_mass}
\end{align}
where $L_\mathrm{LG}$ is given in Appendix~\ref{appendix:C}.
The dimensionless coupling $\lambda_{hk\Phi\Phi}$ is forbidden by the $Z_2$ charge, and $M_{\nu}^{\rm ZB}$ is ignored because the second order of $f$ is negligible unless $s_{2\phi}$ is extremely small.

We assume $g_{e\mu} = g_{e\tau}=0$ to satisfy the LFV bound and fix the value of $g_{ee}$ to give a contribution to the $0\nu 2\beta$ decay by the cocktail diagram.
We can realize such a suitable size of $g_{ee}$ by adjusting the lightest neutrino mass without conflict with the constraints for neutrino masses.
Since Eq.~\eqref{eq:neutrino_mass} is a linear equation for any elements of $g$ and $f$, we can simply solve for the couplings.
We find two trivial solutions:
\begin{enumerate}
\item We assume $f_{\mu\tau}=0$.
Then, we obtain
\begin{align}
g_{ll'}=&\frac{v^2}{r^2s_{2\phi}^2L_\mathrm{LG}m_{hhk}}\left(M_e^{-1}M_\nu M_e^{-1}\right)_{ll'},
\label{eq:solution_g}
\end{align}
for $g_{\mu\mu}$, $g_{\tau\tau}$, and $g_{\mu\tau}$, and
\begin{align}
f_{ll'}=&\frac{16\pi^2 v}{rs_{2\phi}\ln\left(m_{h_1^+}^2/m_{h_2^+}^2\right)\left(m_{l'}^2-m_{l}^2\right)}\left(M_\nu\right)_{ll'},
\label{eq:solution_f}
\end{align}
for $f_{e\mu}$ and $f_{e\tau}$.
When the neutrino mass matrix is given, the couplings are numerically determined.

In Fig.~\ref{fig13}, we show a plot of $\left|G_2\right|/G_F$ as a function of $r^2s_{2\phi}^2$.
The red solid line shows the plot in the case of the Liu-Gu model ($f=0$ in Eq.~\eqref{eq:neutrino_mass}).
The (1,2) element of $M_\nu$ is chosen to be zero (namely $g_{e\mu} = 0$ to satisfy the stringent $\mu\to 3e$ constraint), which means that the neutrino oscillation parameters are related (see Ref.~\cite{Fukuyama:2021iyw}).
The ratio of $g_{e\tau}/g_{\mu\mu}$ is fixed as a solution to reproduce $M_\nu$.
We choose $g_{ee} = 0.3$ and $m_k = 1$ TeV for $r^2s_{2\phi}^2\agt O(1)$.
For $r^2s_{2\phi}^2 \alt O(1)$, it violates the bound of $\tau \to 3e$ process, and therefore, we need to enlarge $m_k$ to satisfy the bound.
Then, the red solid line becomes flat for $r^2s_{2\phi}^2 \alt O(1)$, which can be understood by Eq.~\eqref{tau-to-3e} for fixed $g_{e\tau}/g_{\mu\mu}$.
We note that the $\mu \to e\gamma$ decay can restrict $G_2/G_F$ by Eq.~\eqref{mueg-doubly-charge} for smaller values of $r^2s_{2\phi}^2$.
The blue solid line as the solution of Eq.~\eqref{eq:neutrino_mass} where $g_{ee}=0.3$, $L_\mathrm{LG}=0.1/\left(16\pi^2\right)^2$, and $m_{hhk}=1$\,TeV.
We set $m_k=1$\,TeV for $r^2s_{2\phi}^2 \agt O(0.1)$, while we choose $m_k$ ($> 1$ TeV) to satisfy the $\tau^+ \to \mu^- e^+ e^+$ and $\tau \to 3\mu$ bounds for $r^2s_{2\phi}^2 \agt O(0.1)$.
Since $\left|f_{e\tau}f_{\mu\tau}\right|$ is sufficiently small, the constraint from $\mu\to e\gamma$ does not affect the plot.
In order to obtain the line, we use the best-fit values of the neutrino mass and mixing parameters \cite{Esteban:2020cvm}.
In those solutions on the blue solid line, the $0\nu 2\beta$ contribution from the cocktail diagram can be a little above the current experimental bound if the scalars in the doublets are light as shown in Section~\ref{sec3}.
We find that the model parameters for $r^2s_{2\phi}^2\simeq 0.1$ can induce the Mu-to-$\mubar$ transition with a detectable transition rate just below the current bound shown by the horizontal-dotted line.
\begin{figure}[t]
\center
\includegraphics[width=8cm]{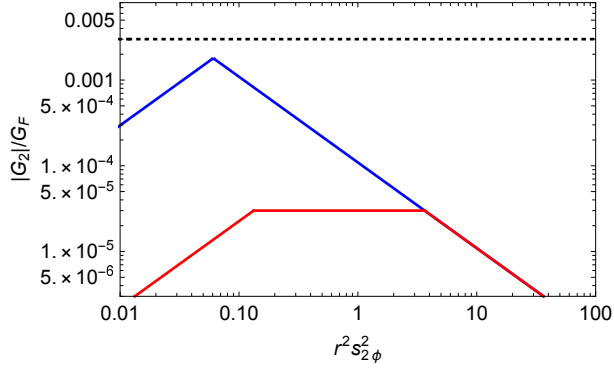}
\caption{
$\left|G_2\right|/G_F$ as a function of $r^2s_{2\phi}^2$.
The red solid line shows the bound in the case where the Zee-Babu contribution is zero (the $\ell \ell h$ coupling is zero).
For $r^2s_{2\phi}^2 < O(1)$, the LFV decay constrains $|G_2|/G_F$.
Even though $g_{\mu\mu}$ becomes larger for smaller $r^2s_{2\phi}^2$, $|G_2|/G_F$ cannot become larger.
The blue solid line is drown if the corresponding parameters to realize the neutrino mass matrix can satisfy the LFV constraints by switching on the $\ell \ell h$ coupling.
The horizontal-dotted line shows the current experimental bound from the Mu-to-$\mubar$ transition.
See the text for more detail on the setup.
}
\label{fig13}
\end{figure}

\item We assume $g_{\mu\tau}=0$.
We obtain the same expressions as Eqs.~\eqref{eq:solution_g} and \eqref{eq:solution_f} for $g_{\mu\mu}$, $g_{\tau\tau}$, $f_{e\mu}$, $f_{e\tau}$, and $f_{\mu\tau}$.
Since the assumption of $g_{\mu\tau}=0$ suppresses the LFV $\tau$ decays, we can more easily evade the constraints from LFV searches.
\end{enumerate}

\section*{Acknowledgements}

This work was supported in part by JSPS KAKENHI Grant Numbers JP18H01210, JP21H00081, JP22H01237 (Y.U.), and JP22K03602 (T.F. and Y.U.).
We thank H.~Sugiyama for making us realize the existence of the one-loop contribution.

\appendix

\section{Scalar spectrum}
\label{appendix:A}

We list the scalar masses
from the scalar potential in
Eq.(\ref{scalar-potential}).

The CP-even neutral scalar mass term is
\begin{equation}
\left(
 \Phi^0 \ \eta^0
\right)
\left(
 \begin{array}{cc}
  v^2 (2\lambda_1 c_\beta^4 + 2\lambda_2 s_\beta^4 +  \lambda_{345} s_{2\beta}^2)
  & 
  - v^2( \lambda_1 c_\beta^2 - \lambda_2 s_\beta^2 - \lambda_{345} c_{2\beta}) \\
  - v^2( \lambda_1 c_\beta^2 - \lambda_2 s_\beta^2 - \lambda_{345} c_{2\beta})
  &
  M^2 + v^2 (\frac12 (\lambda_1 + \lambda_2) - \lambda_{345}) s_{2\beta}^2
 \end{array}
\right)
\left( 
 \begin{array}{c}
 \Phi^0 \\  \eta^0
 \end{array}
\right),
\end{equation}
where $\lambda_{345} = \lambda_3 + \lambda_4 + \lambda_5$
and
\begin{equation}
M^2 \equiv \frac{m_3^2}{c_\beta s_\beta}.
\end{equation}
The mass eigenstates and the mixing angle are conventionally defined as 
\begin{equation}
\left(
 \begin{array}{c}
  H^0 \\ h^0 
 \end{array}
\right) =
\left(
 \begin{array}{cc}
  \cos(\beta-\alpha) & -\sin(\beta-\alpha) \\
  \sin(\beta-\alpha) & \cos(\beta-\alpha)
 \end{array}
\right) 
\left(
 \begin{array}{c}
  \Phi^0 \\ \eta^0 
 \end{array}
\right),
\end{equation}
and the so-called alignment limit ($\cos(\beta-\alpha) \to 0$) without decoupling
is obtained if
\begin{equation}
 \lambda_1 c_\beta^2 - \lambda_2 s_\beta^2 - \lambda_{345} c_{2\beta} \simeq 0.
\end{equation}
The CP-odd neutral scalar mass is
\begin{equation}
m_A^2 = M^2 - 2 \lambda_5 v^2.
\end{equation}

The singly charged scalar mass term is
\begin{equation}
\left( \eta^+ \  h^+
 \right)
 \left(
  \begin{array}{cc}
   m^2_{\eta^+} & v m_{\Phi\Phi h} \\
   v m_{\Phi\Phi h} & m_{h^+}^2
  \end{array}
 \right)
\left( 
 \begin{array}{c}
 \eta^- \\  h^-
 \end{array}
\right),
\end{equation}
where
\begin{equation}
m_{\eta^+}^2 = M^2 - (\lambda_4 + \lambda_5) v^2.
\end{equation}

When the scalar trilinear coupling, such as $m_{hhk}$ and $m_{\Phi\Phi h}$ in Eq.(\ref{scalar-potential}), is large,
 charge symmetry breaking vacua can become a true vacuum 
 as often discussed in supersymmetric models \cite{Frere:1983ag}.
The trilinear coupling, therefore, needs to be bounded to avoid the phase transition to the charge symmetry breaking vacua.
For example,
for a direction of $\langle h^+ \rangle = \langle k^{++} \rangle$,
the value of the potential $\langle V \rangle $ is positive at the charge breaking vacua and 
the charge breaking is not a true vacuum if
\begin{equation}
m_{hhk}^2 < (\lambda_h + \lambda_k + \lambda_{hk} ) ( m_{h^+}^2 + m_k^2 ).
\end{equation}
This is a sufficient condition
since it is acceptable that the electroweak symmetry breaking vacua is metastable in the cosmological time scale.
As a rule of thumb, the trilinear coupling should be smaller than the order of
the quadratic masses.

\section{Loop functions for neutrinoless double beta decay}
\label{appendix:B}

What we need to calculate loop functions is
\begin{equation}
I(m_1,m_2,m_3,m_4) = \int \frac{d^4k}{i(2\pi)^4} \frac{k^\mu k^\nu}{(m_1^2 - k^2) (m_2^2 - k^2) (m_3^2 - k^2) (m_4^2 - k^2)}.
\end{equation}
We obtain
\begin{equation}
I = - \frac{g^{\mu\nu}}{4} \frac{1}{16\pi^2} \frac{1}{m_1^2} F(x,y,z),
\end{equation}
where
\begin{equation}
F(x,y,z) = \frac{x^2 \ln x}{(x-1)(x-y)(x-z)} + \frac{y^2 \ln y}{(y-1)(y-x)(y-z)}  + \frac{z^2 \ln z}{(z-1)(z-x)(z-y)} ,
\end{equation}
and
\begin{equation}
x = \frac{m_2^2}{m_1^2}, \qquad
y = \frac{m_3^2}{m_1^2}, \qquad
z = \frac{m_4^2}{m_1^2}.
\end{equation}

Remarking
\begin{equation}
\frac{1}{m_H^2 - k^2} - \frac{1}{m_A^2 - k^2} =  \frac{-(m_H^2 - m_A^2)}{(m_H^2 - k^2) (m_A^2 - k^2)},
\end{equation}
we obtain the triangle loop part of the cocktail diagram as follows:
\begin{align}
 -(m_H^2 - m_A^2) I (m_1, m_2, m_H,m_A)
=\frac{g^{\mu\nu}}{4} \frac1{16\pi^2}
F_t(a,b,c),
\end{align}
\begin{equation}
F_t(a,b,c) \equiv (b-c) F(a,b,c)
=
\frac{a^2(b-c) \ln a}{(a-1)(a-b)(a-c)} + \frac{b^2 \ln b}{(b-1)(b-a)}  - \frac{c^2 \ln c}{(c-1)(c-a)}  ,
\end{equation}
\begin{equation}
a= \frac{m_2^2}{m_1^2}, \quad b = \frac{m_H^2}{m_1^2}, \quad c = \frac{m_A^2}{m_1^2}.
\end{equation}

The box loop for Fig.\ref{fig5} in Section \ref{sec7} is
\begin{equation}
I(m_1,m_2,m_H,m_N)- I(m_1,m_2,m_A,m_N) = - \frac{g^{\mu\nu}}{4} \frac1{16\pi^2} \frac{1}{m_1^2}
F_b (a,b,c,d),
\end{equation}
\begin{equation}
F_b (a,b,c,d) \equiv F(a,b,d) - F(a,c,d),
\end{equation}
where 
\begin{equation}
d = \frac{m_N^2}{m_1^2}.
\end{equation}

\section{Loop functions for neutrino masses}
\label{appendix:C}

\begin{figure}[t]
\center
\includegraphics[width=8.2cm]{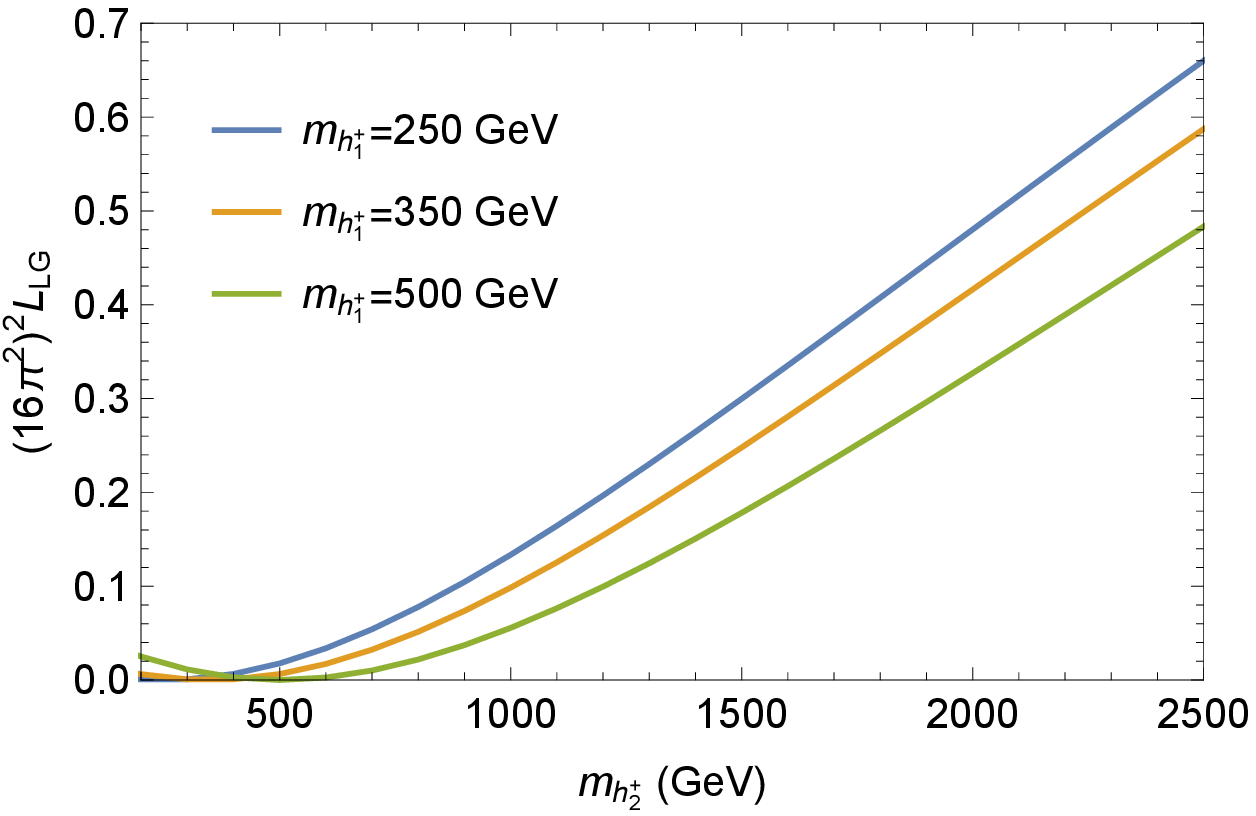}
\includegraphics[width=8cm]{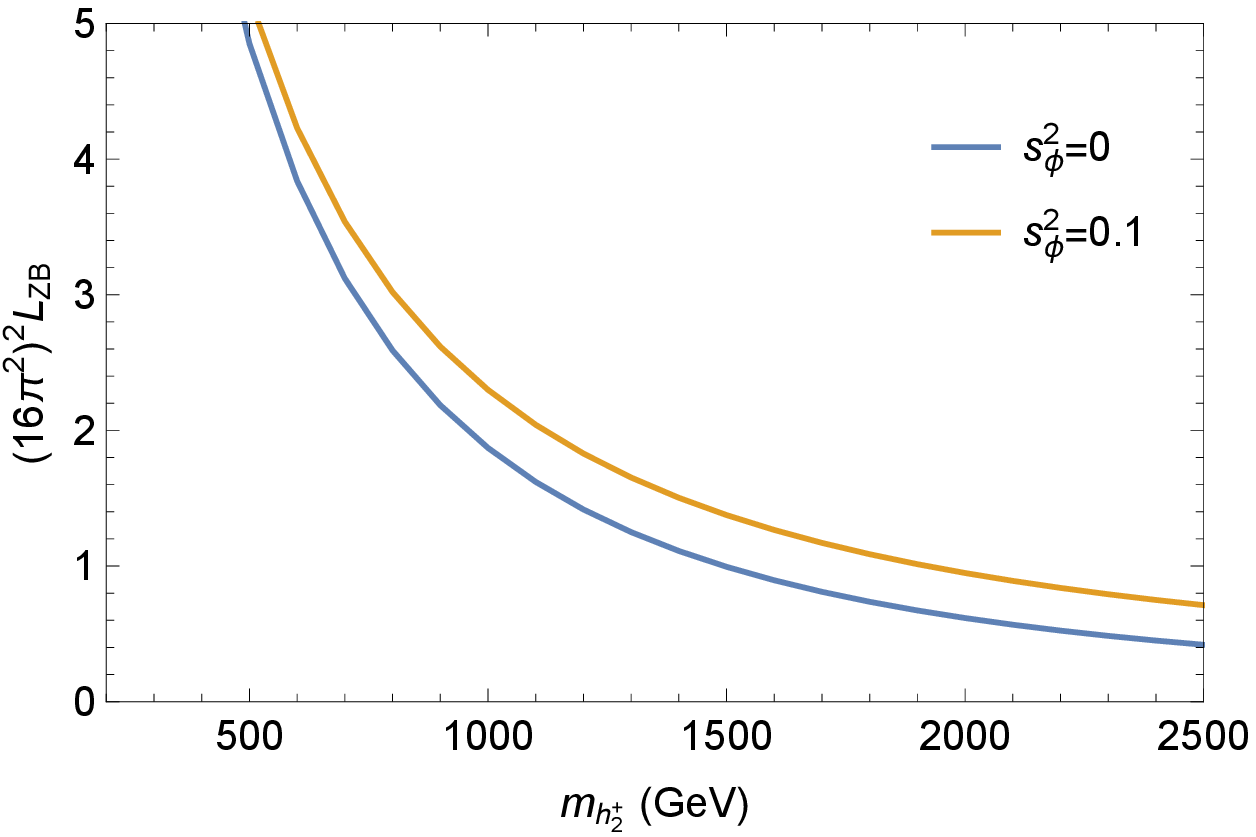}
\caption{
Plots of $(16 \pi^2)^2 L_{\rm LG}$ (left) and $(16 \pi^2)^2 L_{\rm ZB}$ (right).
We choose $m_k = 1$ TeV.
For $L_{\rm ZB}$, we choose $m_{h_1^+} = 300$ GeV. 
}
\label{fig:loop}
\end{figure}

The loop integrals for the neutrino masses induced by the diagrams in Figs.\ref{fig2} and \ref{fig3}
are
\begin{align}
L_1 (x,y) {\bf 1} &= \int \frac{d^4 k_1}{i(2\pi)^4} \int \frac{d^4 k_2}{i(2\pi)^4}
\frac{k_1\!\!\!\!\!/  \ k_2\!\!\!\!\!/}{(1-(k_1+k_2)^2) k_1^2 k_2^2 (x - k_1^2) (y - k_2^2)}, \\
L_2 (x,y)&= \int \frac{d^4 k_1}{i(2\pi)^4} \int \frac{d^4 k_2}{i(2\pi)^4}
\frac{1}{(1-(k_1+k_2)^2) k_1^2 k_2^2 (x - k_1^2) (y - k_2^2)}.
\end{align}
We define loop functions for the respective diagrams as
\begin{align}
L_{\rm LG} &= L_1 (x,x) + L_1 (y,y) - 2 L_1(x,y), \\
L_{\rm ZB} &= s_\phi^4 L_2 (x,x) + c_\phi^4 L_2 (y,y) + 2 c_\phi^2 s_\phi^2 L_2 (x,y).
\end{align}
In Fig.\ref{fig:loop}, we plot the loop functions for
\begin{equation}
x = \frac{m_{h_1^+}^2}{m_k^2}, \qquad y = \frac{m_{h_2^+}^2}{m_k^2}.
\end{equation}

\section{LFV constraints for $f$ and $g$ coupling matrices}
\label{appendix:D}

The couplings $f$ and $g$ in Eq.(\ref{f-and-g})
can induce LFV decay processes.
Let us summarize the constraints from the processes.

The branching ratio of the $\mu \to e \gamma$ decay induced by the $f$ coupling is
\begin{equation}
{\rm Br} (\mu \to e\gamma) = \frac{3 \alpha}{16 \pi} \left| \frac{f_{e\tau} f_{\mu\tau}^*}{3 G_F m_{h^+}^2} \right|^2.
\end{equation}
The experimental bound of the branching ratio is \cite{Adam:2013mnn}
\begin{equation}
{\rm Br}(\mu \to e \gamma) < 4.2 \times 10^{-13},
\end{equation}
and the $f_{e\tau} f_{\mu\tau}^*$ product is constrained as
\begin{equation}
| f_{e\tau} f^*_{\mu\tau}| \alt 1.0 \times 10^{-3} \times \frac{m_{h^+}^2}{(1\, {\rm TeV})^2}.
\label{fetau-fmutau}
\end{equation}
The branching ratio of the decay induced by the $g$ coupling is
\begin{equation}
{\rm Br} (\mu \to e\gamma) = \frac{3\alpha}{16\pi} \left| 
\frac{4(g_{ee} g^*_{\mu e} + g_{e\mu} g^*_{\mu \mu} + g_{e\tau} g^*_{\mu \tau}) }{3 G_F m_k^2} 
\right|^2,
\end{equation}
and the experimental bound of the coupling can be similarly written.
Since the $g$ coupling can also induce the Mu-to-$\mubar$ transition operator in Eq.(\ref{Lag:G2}),
 it is convenient to express the bound as follows:
\begin{equation}
\frac{|G_2|}{G_F} < 3.8 \times 10^{-6} \left| \frac{g_{ee} g_{\mu\mu} }{g_{ee} g_{\mu e}^*+g_{e\mu} g_{\mu\mu}^*+ g_{e\tau} g_{\mu\tau}^* } \right|.
\label{mueg-doubly-charge}
\end{equation}
The $g$ coupling can induce the $\mu$\dash$e$ conversion in nuclei \cite{SINDRUMII:2006dvw}
with a large $\ln (m_\mu^2/m_k^2)$ correction~\cite{Raidal:1997hq}.
The current experiment gives a weaker constraint than the $\mu \to e\gamma$ bound.

The $\tau \to \mu (e) \gamma$ decays \cite{Zyla:2020zbs},
\begin{equation}
\{ {\rm Br}(\tau \to e\gamma), {\rm Br}(\tau \to \mu\gamma) 
\}
< \{ 3.3,4.4 \}\times 10^{-8} , \\
\end{equation}
can also give bounds to the $f$ and $g$ couplings, e.g.,
\begin{equation}
\{ | f_{e\mu} f^*_{\tau\mu} |, | f_{\mu e} f^*_{\tau e}|
\} \alt \{ 0.66, 0.76 \} \times \frac{m_{h^+}^2}{(1\, {\rm TeV})^2},
\end{equation}
though 
the bounds are not stringent.

\begin{figure}[t]
\center
\includegraphics[width=8cm]{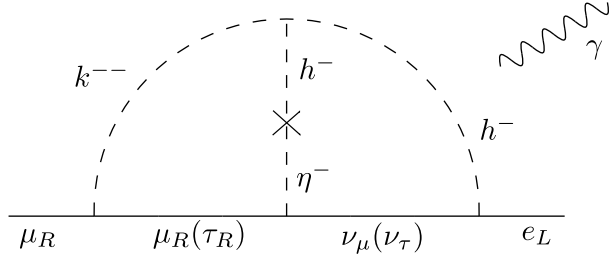}
\caption{
Two-loop diagram which can generate the $\mu \to e \gamma $ decay
in the model with the $h^-$\dash$\eta^-$ mixing in the $Z_2$-charge B.
The photon line can be attached to the lines of the charged particles.
}
\label{fig4}
\end{figure}

We comment that
the $h^+$\dash$\eta^+$ mixing does not induce a new one-loop bound from
the $\mu \to e\gamma$ decay, since an active neutrino mass is inserted in the loop diagram
with the $h^+$\dash$\eta^+$ mixing.
In the two-loop level, there can be a diagram given in Fig.\ref{fig4}.
We can verify that the two-loop contribution does not provide a significant constraint
as long as the $f$ couplings satisfy the one-loop level constraint
and the $g_{\mu \tau}$ coupling is suppressed rather than $g_{\mu\mu}$ to reproduce the neutrino
mass matrix.

The $g$ coupling can generate 3-body LFV decays as follows:
\begin{equation}
{\rm Br} (l_a^- \to l_b^+ l_c^- l_d^-) 
= \frac{1}{2(1+\delta_{cd})}  \left| 
\frac{g_{ab} g_{cd}^*}{G_F m_k^2}
 \right|^2 \times {\rm Br} (l_a^- \to l_b^-  \nu\bar\nu) .
 \label{3-body-LFV}
\end{equation}
The bound of the $\mu \to 3e$ decay process \cite{Bellgardt:1987du},
\begin{equation}
 {\rm Br}(\mu \to 3 e) <1.0 \times 10^{-12},
\end{equation}
gives the most stringent constraint to the Mu-to-$\mubar$ transition in the model.
We find 
\begin{equation}
\frac{|G_2|}{G_F} = \sqrt{{\rm Br} (\mu \to 3e)}
\frac1{2\sqrt2} \left| \frac{g_{\mu \mu}}{g_{e\mu}} \right| \alt 3.5 \times 10^{-7} \left| \frac{g_{\mu \mu}}{g_{e\mu}} \right| .
\end{equation}
Let us calculate the numerical upper bounds of $|G_2|/G_F$ 
allowed by the constraints from the LFV decays \cite{Zyla:2020zbs}:
\begin{eqnarray}
&&\{{\rm Br}(\tau \to 3e), {\rm Br}(\tau \to 3\mu) 
\}
< \{2.7,2.1 
 \}\times 10^{-8} , 
 \label{bound:tau-decay-1}
 \\
%
&&\{
{\rm Br}(\tau^- \to e^+ \mu^- \mu^-),
{\rm Br} (\tau^- \to \mu^+ e^- e^-) \}
< \{
1.7,    
1.5  
\}\times 10^{-8} .
\label{bound:tau-decay-2}
\end{eqnarray}
The LFV decay bounds can give the upper bounds of the Mu-to-$\mubar$ transition as
\begin{eqnarray}
\frac{|G_2|}{G_F}& \simeq& \sqrt{   \frac{{\rm Br} (\tau \to 3e)}{{\rm Br} (\tau \to e \nu\bar\nu)}     }
\frac1{2\sqrt2} \left| \frac{g_{\mu \mu}}{g_{e\tau}} \right| \alt 1.4 \times 10^{-4} \left| \frac{g_{\mu \mu}}{g_{e\tau}} \right| ,
\label{tau-to-3e} \\
\frac{|G_2|}{G_F}& \simeq& \sqrt{   \frac{{\rm Br} (\tau^- \to \mu^+ e^- e^-)}{{\rm Br} (\tau \to \mu \nu\bar\nu)}     }
\frac1{2\sqrt2} \left| \frac{g_{\mu \mu}}{g_{\mu\tau}} \right| \alt 1.0 \times 10^{-4} \left| \frac{g_{\mu \mu}}{g_{\mu\tau}} \right| , 
\label{tau-to-muee}\\
\frac{|G_2|}{G_F}& \simeq& \sqrt{   \frac{{\rm Br} (\tau^- \to e^+ \mu^-\mu^-)}{{\rm Br} (\tau \to e \nu\bar\nu)}     }
\frac1{2\sqrt2} \left| \frac{g_{e e}}{g_{e\tau}} \right| \alt 1.1 \times 10^{-4} \left| \frac{g_{ee}}{g_{e\tau}} \right| , \\
\frac{|G_2|}{G_F}& \simeq& \sqrt{   \frac{{\rm Br} (\tau \to 3\mu)}{{\rm Br} (\tau \to \mu \nu\bar\nu)}     }
\frac1{2\sqrt2} \left| \frac{g_{ee}}{g_{\mu\tau}} \right| \alt 1.2 \times 10^{-4} \left| \frac{g_{ee}}{g_{\mu\tau}} \right| .
\label{tau-to-3mu}
\end{eqnarray}

\section{Scalar spectrum in the minimal left-right model}
\label{appendix:E}

The scalar potential in terms of a bidoublet $\Phi_b : ({\bf 2}, {\bf 2}, 0)$
and a triplet $\Delta : ({\bf 1}, {\bf 3}, 1)$
under $(SU(2)_L, SU(2)_R, (B-L)/2)$ is
\begin{align}
V &= 
m^2_\Delta {\rm tr} \Delta \Delta^\dagger
+ m^2_\Phi {\rm tr} \Phi_b \Phi_b^\dagger
+ \mu^2_\Phi ({\rm tr} \Phi_b \Phi_b + {\rm H.c.}) \\
&+
\kappa_1 ({\rm tr} \Delta \Delta^\dagger)^2 + \kappa_2 ({\rm tr} \Delta \Delta)({\rm tr} \Delta^\dagger \Delta^\dagger) \nonumber\\
&+
\kappa_3 ({\rm tr} \Phi_b \Phi_b^\dagger)^2 + \kappa_4  ({\rm tr} \Phi_b \Phi_b)  ({\rm tr} \tilde\Phi_b \tilde\Phi_b) 
+ \kappa_5 ( ({\rm tr} \Phi_b \Phi_b)^2 +  {\rm H.c.}) \nonumber\\
&
+ \kappa_6 ({\rm tr} \Phi_b \Phi_b^\dagger) ( {\rm tr} \Phi_b \Phi_b +  {\rm H.c.}) \nonumber\\
&+
\kappa_7 ({\rm tr} \Phi_b \Phi_b^\dagger)({\rm tr} \Delta \Delta^\dagger)
+ 
\kappa_8 ({\rm tr} \Delta \Delta^\dagger) ( {\rm tr} \Phi_b \Phi_b +  {\rm H.c.}) 
+
\kappa_9{\rm tr} \Phi_b^\dagger \Phi_b \Delta \Delta^\dagger, \nonumber
\end{align}
where
\begin{equation}
\tilde \Phi_b = \epsilon \Phi_b^* \epsilon^T.
\end{equation}
Denoting 
\begin{equation}
\Phi_b = 
\left(
 \begin{array}{cc}
 \phi^+_u & \phi_d^0 \\
 \phi^0_u & \phi_d^-
 \end{array}
\right),
\end{equation}
we obtain the mass term of the
singly charged scalars as
\begin{equation}
\left( \begin{array}{ccc}
 \Delta^- & \phi_u^- & \phi_d^-
\end{array}
\right) 
{\cal M}^2_+
\left( 
\begin{array}{c}
 \Delta^+ \\ \phi_u^+ \\ \phi_d^+
\end{array}
\right) ,
\end{equation}
\begin{equation}
{\cal M}^2_+ =  \frac12 \kappa_9 
\left(
 \begin{array}{ccc}
  v^2 \cos2\beta & - \sqrt2 v v_R \cos\beta & - \sqrt2 v v_R \sin\beta \\
   - \sqrt2 v v_R \cos\beta & 2 v_R^2 \cos^2 \beta \sec2\beta &  v_R^2 \tan2\beta \\
   - \sqrt2 v v_R \sin\beta & v_R^2 \tan2\beta & 2 v_R^2 \sin^2\beta \sec2\beta
 \end{array}
\right),
\end{equation}
where $v_R = \langle \Delta^0 \rangle$,
$v= \sqrt{ \langle \phi_u^0 \rangle^2 + \langle \phi_d^0 \rangle^2}$ and $\tan\beta = \langle \phi_u^0 \rangle/ \langle \phi_d^0 \rangle$.
We have used minimization conditions of the scalar potential.
The mass eigenstates are obtained by
\begin{equation}
\left(
 \begin{array}{c}
  \omega^+_R \\ H^+ \\ \omega_L^+
 \end{array}
\right)
=
\left(
 \begin{array}{ccc}
   c_\phi & s_\phi & 0 \\
   -s_\phi & c_\phi & 0 \\
   0 & 0 & 1 
 \end{array}
\right)
\left(
 \begin{array}{ccc}
   1 & 0 & 0 \\
   0 & c_\beta & s_\beta  \\
   0 & -s_\beta & c_\beta   
 \end{array}
\right)
\left(
 \begin{array}{c}
  \Delta^+ \\ \phi_u^+ \\ \phi_d^+
 \end{array}
\right),
\end{equation}
where
\begin{equation}
\tan\phi = -\frac{v}{v_R} \frac{\cos2\beta}{\sqrt2}.
\end{equation}
The $\omega_L^+$ and $\omega_R^+$ bosons are massless and are eaten by $W_L^+$ and $W_R^+$.
The physical charged scalar mass is
\begin{equation}
M_{H^+}^2 = \kappa_9 (v_R^2 \sec2\beta +\frac12 v^2 \cos2\beta).
\end{equation}
We need $\kappa_9 <0$ for $v_u > v_d$ (i.e., $\cos2\beta <0$).
We note that the $\Delta^{--}$\dash$H^+$\dash$H^+$ term is calculated as
\begin{align}
V \supset
\Delta^{--} H^+ H^+ ( 2 \kappa_2 v_R \sin^2\phi - \frac{\kappa_9}{2\sqrt2} v \sin2\phi).  
\end{align}

The doubly charged scalar mass is obtained as
\begin{equation}
M_{\Delta^{++}}^2  = 4 \kappa_2 v_R^2 + \kappa_9 v^2 \cos2\beta.
\end{equation}

For neutral components,
${\rm Im} \Delta^0$ and ${\rm Im} (- s_\beta \phi_{u}^0 + c_\beta \phi_{d}^0) $ are 
eaten by two neutral gauge bosons.
The CP-odd scalar mass is
\begin{equation}
M_A^2 = \kappa_9 v_R^2 \sec 2\beta + 4 (\kappa_4 - 2 \kappa_5) v^2.
\end{equation}
The mass matrix of three CP-even neutral scalars is 
\begin{equation}
\left(
 \begin{array}{ccc}
  M_{hh}^2 & M^2_{hH} & M^2_{h\Delta} \\
   M^2_{hH} & M^2_{HH} & M^2_{H\Delta} \\
     M^2_{h\Delta}   &    M^2_{H\Delta}        & M^2_{\Delta \Delta}
 \end{array}
\right),
\end{equation}
\begin{align}
M_{hh}^2 &= 4 v^2 ( \kappa_3 + (\kappa_4 +2\kappa_5) \sin^2 2\beta - 4\kappa_6 \sin2\beta), \\
M_{hH}^2 &= -4 v^2 ( \kappa_6 - (\kappa_4 +2\kappa_5) \sin 2\beta ) \cos2\beta, \\
M_{HH}^2 & = \kappa_9 v_R^2 \sec 2\beta  + 4 (\kappa_4 + 2\kappa_5) v^2 \cos^22\beta, \\
M_{h\Delta}^2 &= 2v v_R (\kappa_7 -2 \kappa_8 \sin2\beta + \kappa_9 \sin^2\beta),\\
M_{H\Delta}^2 &= v v_R (-4 \kappa_8 \cos2\beta + \kappa_9 \sin2\beta), \\
M_{\Delta\Delta}^2 &= 4 \kappa_1 v_R^2.
\end{align}

\end{document}